\documentclass[twocol]{ametsocV6.1}

\usepackage[utf8]{inputenc}
\usepackage{amsmath, amsfonts, amssymb}
\usepackage{bm}
\usepackage[bb=dsserif]{mathalpha}
\usepackage{anyfontsize}

\usepackage{xcolor}

\usepackage{url}

\usepackage{graphicx}
\usepackage{caption}
\usepackage{subcaption}
\newcommand{\pointA}{$14^\circ$N, $145^\circ$E}
\newcommand{\pointB}{$11^\circ$S, $60^\circ$E}

\title{Uncertainty-permitting machine learning reveals sources of dynamic sea level predictability across daily-to-seasonal timescales}

\authors{
Andrew Brettin\aff{a}\correspondingauthor{Andrew Brettin, brettin@cims.nyu.edu},
Laure Zanna\aff{a},
Elizabeth A. Barnes\aff{b}
}

\affiliation{\aff{a}{Courant Institute of Mathematical Sciences, New York University, New York, New York}, \aff{b}{Department of Atmospheric Science, Colorado State University, Fort Collins, Colorado}}

\abstract{Reliable dynamic sea level forecasts are hindered by numerous sources of uncertainty on daily-to-seasonal timescales (1--180 days) due to atmospheric boundary conditions and internal ocean variability. Studies have demonstrated that certain initial states can extend predictability horizons; thus, identifying these initial conditions may help improve forecast skill. 
Here, we identify sources of dynamic sea level predictability on daily-to-seasonal timescales using neural networks trained on CESM2 large ensemble data to forecast dynamic sea level. The forecasts yield not only a point estimate for sea level but also a standard deviation to quantify forecast uncertainty based on the initial conditions.
Forecasted uncertainties can be leveraged to identify state-dependent sources of predictability at most locations and forecast leads. Network forecasts, particularly in the low-latitude Indo-Pacific, exhibit skillful deterministic predictions and skillfully forecast exceedance probabilities relative to local linear baselines. 
For networks trained at Guam and in the western Indian Ocean, the transfer of sources of predictability from local sources to remote sources is presented by the deteriorating utility of initial condition information for predicting exceedance events. Propagating Rossby waves are identified as a potential source of predictability for dynamic sea level at Guam. In the Indian Ocean, persistence of thermosteric sea level anomalies from the Indian Ocean Dipole may be a source of predictability on subseasonal timescales, but El Ni\~no drives predictability on seasonal timescales. This work shows how uncertainty-quantifying machine learning can help identify changes in sources of state-dependent predictability over a range of forecast leads.
}
\begin{document}

\maketitle

%
%
%
\newpage 
\statement

Uncertainty-quantifying neural networks trained to forecast dynamic sea level anomalies on daily-to-seasonal timescales skillfully identify sources of state-dependent predictability and forecast exceedance probabilities. We use these predicted probabilities to identify how sources of state-dependent predictability change over daily-to-seasonal timescales. At Guam, Rossby waves are a source of dynamic sea level predictability. In the western Indian Ocean, El Ni\~no emerges as a source of predictability, possibly by acting as a precursor to the Indian Ocean Dipole. This work may clarify how sources of predictability change between daily and seasonal timescales, which may help improve forecasts at these lead times.

%
%
%


\section{Introduction}
\label{sec:introduction}
Dynamic sea level, defined as the height of the sea surface above the geoid (excluding the inverse-barometric imprint from atmospheric loading), is modulated by a variety of processes from the atmosphere and ocean \citep{gregory2019concepts, griffies2016omip}. Dynamic sea level responds to atmospheric forcing through local processes, such as surface buoyancy fluxes \citep{hochet2024advection,cabanes2006contributions, gill1973theory} and Ekman pumping \citep{qu2022drivers,piecuch2011mechanisms} and through remote processes such as time-dependent Sverdrup dynamics \citep{chen2023topography, qiu2002large} and baroclinic response to wind stress \citep{cabanes2006contributions}. Intrinsic ocean variability impacts dynamic sea level by mass redistribution \citep{fukumori1998nature}, advection and diffusion of steric anomalies \citep{hochet2024advection} and barotropic and baroclinic instabilities \citep{penduff2010impact}.


The wide variety of processes impacting dynamic sea level can cause significant variations in sea level over a range of timescales with potentially adverse consequences. For instance, variations in dynamic sea level associated with the El Ni\~no-Southern Oscillation (ENSO) can be as large as 20-30cm \citep{becker2012sea_a}, comparable to the observed increase in global mean sea level due to contemporary anthropogenic climate change over the past century \citep{frederikse2020causes}. Drops in sea level in the western Pacific associated with El Ni\~no can expose shallow reefs, leading to coral damage \citep{widlansky2014interhemispheric}. As another example, sea level variability can escalate the risk of high-tide flooding, resulting in disruptions such as salinated aquifers \citep{sukop2018high, becker2012sea_b}, impediments to transportation and commercial activity \citep{hino2019high}, and damages to wastewater treatment facilities \citep{hummel2018sea}. Global mean sea level rise has intensified the frequency of such nuisance ``fair-weather” floods \citep{li2022contributions, hino2019high}, and projections suggest that high-tide flooding events will become more common in many locations in the future \citep{sweet2014extreme}.

The impacts of dynamic sea level fluctuations have motivated a variety of efforts to improve forecasting of sea level variability in order to help mitigate such effects. Approaches to forecasting have included statistical techniques, such as canonical correlation analysis \citep{chowdhury2015sea}, multivariate linear regression \citep{widlansky2017multimodel}, and probabilistic models \citep{dusek2022novel}. Furthermore, contemporary developments in coupled dynamical models have aided their ability to generate skillful forecasts of sea level on seasonal outlooks \citep{miles2014seasonal, widlansky2023quantifying, balmaseda2024skill}.

Dynamic sea level variability on seasonal-to-interannual timescales \citep[1-24 months, ][]{jacox2020seasonal} has received considerable attention, and numerous studies have used observations and simulations to investigate the predictability of dynamic sea level on these time horizons \citep{balmaseda2024skill, wang2023seasonal, doi2020skill, fraser2019investigating, miles2014seasonal}. The interest in seasonal-to-interannual dynamic sea level variability may be driven, in part, by significant relationships with indices of climate variability on these timescales, such as the El Ni\~no-Southern Oscillation, the Indian Ocean Dipole, the Southern Annular Mode and the North Atlantic Oscillation \citep{roberts2016drivers, chowdhury2007enso, miles2014seasonal, aparna2012signatures, kenigson2018decadal}. 
Studies have also highlighted the potential utility of sea level forecasts on subseasonal-to-seasonal timescales \citep[15-60 days, ][]{demott2021benefits, amaya2022subseasonal, arcodia2024subseasonal}. This forecasting horizon is a critical time window for municipalities to take preemptive action to mitigate damage from high-tide flooding. However, forecasts on subseasonal-to-seasonal timescales have generally been regarded as a challenging timescale for prediction in the earth system \citep{vitart2017subseasonal, mariotti2018progress, nas2016next}. While the slow internal variability of the ocean compared to the atmosphere can provide some predictability, memory from the initial conditions of the atmosphere is typically lost beyond timescales of two weeks \citep{krishnamurthy2019predictability, lorenz1969predictability}. Forecasting techniques such as ocean-dynamic persistence, in which dynamical ocean models are forced by climatological atmospheric conditions, have shown some success, though chaotic atmospheric dynamics can still occlude the predictability conferred by ocean conditions and pose inherent forecasting challenges \citep{feng2025indications}.

Recent approaches that have allowed for the development of useful forecasts of geophysical conditions on subseasonal-to-seasonal timescales have focused on identifying specific initial conditions that can result in more skillful forecasts \citep{mariotti2020windows}. Studies have established that certain initial conditions from the atmosphere or ocean can provide more predictability for the dynamics of geophysical fields than others \citep{christensen2020value, frame2013flow, kalnay1987forecasting}.  Thus, identifying such state-dependent sources of predictability can enable forecasts to be made on lead times that would normally not be considered \citep{albers2019priori}. Understanding sources of state-dependent predictability---and how these sources vary by forecast lead---can help bridge the gap between daily and seasonal forecasts.

Machine learning approaches, such as artificial neural networks (ANN), can help identify sources of state-dependent predictability directly from data. \citet{mayer2021subseasonal} showed how a classification artificial neural network trained to predict the sign of geopotential height anomalies could be used for a priori identification of skillful forecasts. In particular, activation functions can be leveraged to output class probabilities targeting predictable outcomes and their associated initial conditions. As an alternative method for estimating forecast uncertainty, \citet{gordon2022incorporating} explored state-dependent predictability of sea surface temperatures on decadal timescales using a regression neural network trained on a Gaussian maximum-likelihood based loss function. The regression networks yield stochastic uncertainty estimates for the prediction which can be directly applied to identify modes of variability associated with climate predictability, such as the Atlantic Meridional Variability and Interdecadal Pacific Oscillation. 

Here, we apply a similar approach to \citet{gordon2022incorporating} to investigate state-dependent sources of predictability of dynamic sea level anomalies on daily-to-seasonal timescales (1--180 days). We train regression neural networks to make probabilistic forecasts of sea level using simulated fields from the Community Earth System Model, version 2 Large Ensemble project \citep{danabasoglu2020community, rodgers2021ubiquity}. We examine how sources of state-dependent predictability change over daily-to-seasonal timescales, and identify these sources over different forecast leads.

\section{Methods}
\label{sec:methods}
\subsection{Data}
\label{subsec:data}
We use simulated fields from the Community Earth System Model, version 2 (CESM2) Large Ensemble (LENS2) dataset \citep{danabasoglu2020community, rodgers2021ubiquity}. CESM2 is a fully-coupled earth system model run using a $1^\circ$ nominal horizontal resolution in the ocean and atmosphere. The atmosphere is simulated using the finite-volume dynamical core of the Community Atmosphere Model, version 6 \citep[CAM6,][]{lin1997explicit} and the ocean is simulated by solving the primitive equations employing the hydrostatic approximation using the Parallel Ocean Program, version 2 \citep[POP2,][]{smith2010parallel}. The data is from the 250-year simulation period 1850--2100, with radiative forcing prescribed by the historical record from 1850--2015 and by the CMIP6 SSP370 forcing scenario from 2016--2100 \citep{oneill2016cmip6}. Ensemble simulations are initialized from a combination of different ``macro-perturbations" of a preindustrial simulation state (based on different AMOC phases) and ``micro-perturbations" (by adding minuscule noise to surface air temperature fields), as detailed in \citet{rodgers2021ubiquity}. In addition to the different initialization procedures for different ensemble members in the LENS2 project, a further distinction between sets of ensemble members is made by imposing two different diagnostic surface forcing fields from biomass burning: in 50 of the 100 ensemble members, the CMIP6 protocol is followed; in the remaining 50 ensemble members, an 11-year running mean filter is applied to smooth large interannual variability in the forcing fields. We select nine of the smoothed biomass-burning ensemble members for this analysis, using three micro-perturbations for each of the three macro-perturbations that were available. Using different ensemble members allows us to identify climate drivers of sea level predictability which are robust under different climate forcings and internal variability. Because the ocean is simulated on a displaced-dipole grid, we regrid all oceanic variables to the uniform spherical-coordinate atmospheric grid using bilinear interpolation prior to any analysis. While a conservative interpolation strategy could also be used, we justify the simpler bilinear interpolation technique as sufficient for the prediction of deseasonalized quantities.

We generate uncertainty-permitting forecasts for 5-day averaged dynamic sea level anomalies on the $1^\circ$ grid at various daily-to-seasonal forecasting leads. For CMIP6 model evaluations \citep{eyring2016overview, ipcc2021ocean}, dynamic sea level is defined by the deviation of the sea surface height from the global mean at that timestep, excluding the inverse-barometer contribution to sea surface height from atmospheric pressure loading \citep{griffies2014assessment, gregory2019concepts, wunsch1997atmospheric}. In CESM2, it is computed using the implicit free-surface formulation of the barotropic equations from \citet{dukowicz1994implicit}  \citep{fasullo2020sea, smith2010parallel}.

As model inputs, we use 5-day averaged dynamic sea level \citep[ZOS,][]{griffies2016omip}, sea surface temperatures (SST), and surface zonal and meridional wind fields (UAS and VAS, respectively) from $60^\circ$S to $60^\circ$N. These input fields are coarsened to $5^\circ$ resolution to reduce the input dimensionality and help with understanding large-scale drivers, yielding a feature vector of 6,014 inputs.

To obtain anomalies, we detrend and deseasonalize all variables. We use detrended variables to focus on the internal variability of the system, due to the challenges of learning out-of-sample relationships with neural networks. Variables are detrended using a locally-fitted fifth-order polynomial computed over the full 250-year period for each ensemble member. The seasonal cycle is similarly removed by subtracting climatological daily averages at each grid point. Of the nine ensemble members used, seven are used for training (128,233 samples), and one is used for validation and testing, respectively (18,319 samples each). Values are standardized in time as a preprocessing step prior to training \citep{lecun2002efficient}. That is, for each location and field variable, the mean and standard deviation are taken over all samples in the training dataset and used to standardize the training, validation, and testing set.

\subsection{Machine learning framework}
\label{subsec:ml_framework}

\begin{figure*}[!ht]
    \centering
    \includegraphics[width=\textwidth]{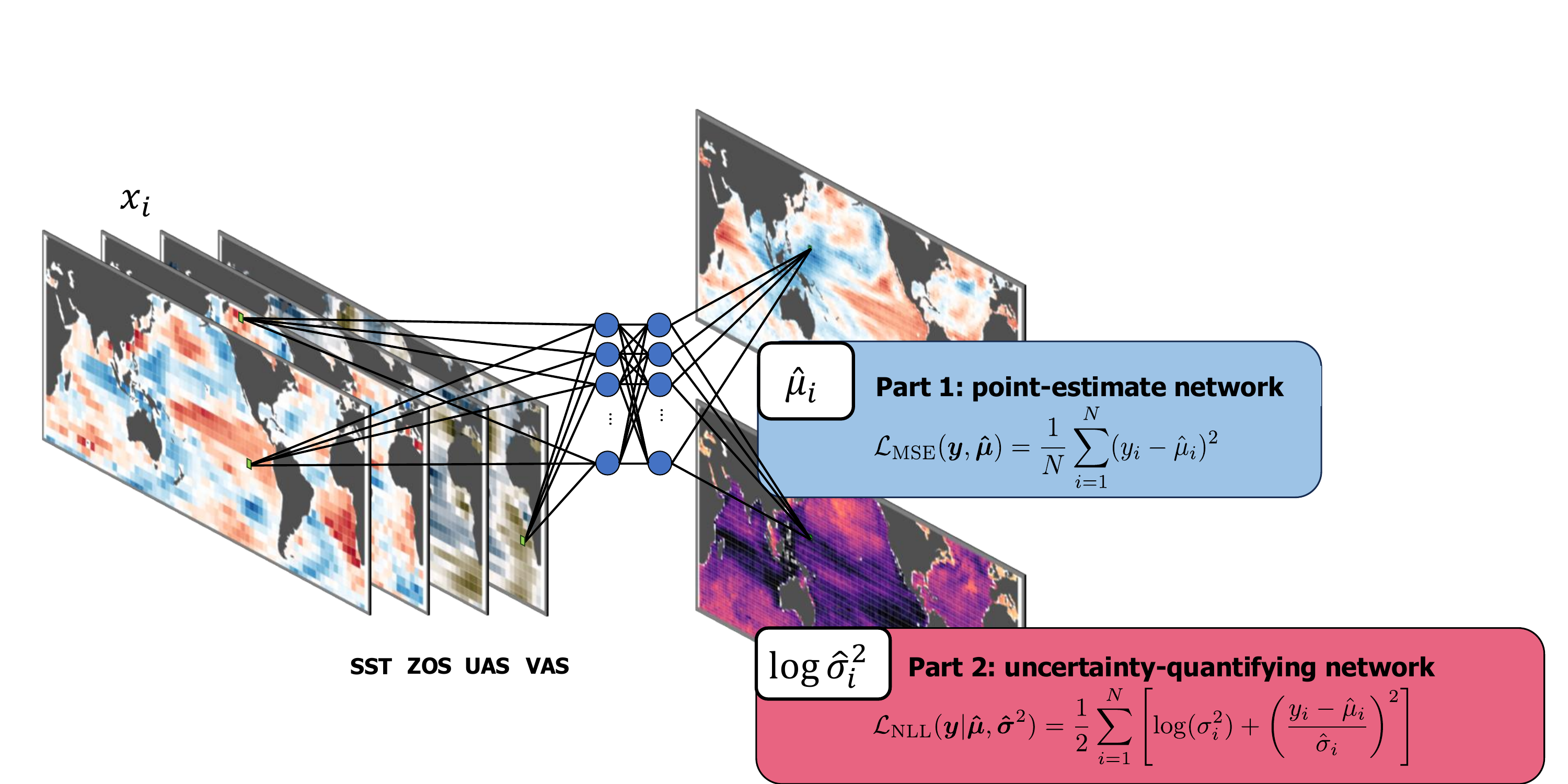}
    \caption{Network schematic. First, the point-estimate network consumes the  input vector $x_i$ of $5^\circ$-coarsened sea surface temperatures, dynamic sea levels, and zonal and meridional winds and forecasts a point-estimate for dynamic sea level for a specified forecast lead at a given location on the $1^\circ$ grid. This network is trained on the Mean Squared Error. Then, the uncertainty-quantifying network uses the same input fields to quantify the forecast uncertainty for a specified forecast lead at a given location on the $1^\circ$ grid. This network is trained using the Gaussian negative log-likelihood loss. To obtain spatial coverage, separate networks are trained at 6,590 different gridpoints on the $1^\circ$ grid for each forecast lead.}
    \label{fig:network_architecture}
\end{figure*}

For each prediction location and forecast lead, we train two fully-connected regression artificial neural networks (ANN) to forecast dynamic sea level: one network predicts a point estimate for the forecast, and the second quantifies the uncertainty associated with the point estimate. Figure~\ref{fig:network_architecture} illustrates the set-up of each network. Each network consumes the same 6,014-dimensional vector of SST, ZOS, UAS, and VAS at every $5^\circ$ gridpoint as input, and issues predictions at a single location on the $1^\circ$ grid specified by the training data for that location. To obtain spatial coverage, we train networks at every other gridpoint latitude and longitude between $60^\circ$S and $60^\circ$N (6,590 locations) for time lags of 10, 20, 60, and 120 days. Although it is in principle possible to train a single network with outputs at every gridpoint instead, training separate networks for each location allows for better predictions at each individual location. Each network is trained for less than 15 minutes using Pytorch \citep{paszke2019pytorch} on 1 Intel Xeon Gold 6240 processor with 16 GB memory.

Given an input $x_i \in \mathbb{R}^{6,014}$ and model parameters $\boldsymbol{\theta}$, the first network outputs a prediction $\hat{\mu}_i = \hat{\mu}(x_i| \boldsymbol{\theta})$ for the target variable $y_i \in \mathbb{R}$. During training, parameters of the model are adjusted to optimize the Mean Square Error (MSE) loss function of the predicted values $\boldsymbol{\hat{\mu}} = (\hat{\mu}_1, \dots, \hat{\mu}_N)$ given the targets $\boldsymbol{y} =(y_1, \dots, y_N)$:
\begin{equation}
    \mathcal{L}_{\textrm{MSE}}(\boldsymbol{y}, \boldsymbol{\hat{\mu}}) = \frac{1}{N} \sum_{i=1}^N (y_i - \hat{\mu}_i)^2.
    \label{eq:mse_loss}
\end{equation} 

The point estimate networks contain two hidden layers of 10 nodes each. Hidden layer nodes are equipped with rectified linear unit (ReLU) activation functions. Because the feature vector is relatively large, we apply dropout regularization \citep{srivastava2014dropout} with a dropout probability of 0.1 between the input layer and the first hidden layer. Parameters are updated using the Adam optimizer \citep{kingma2014adam} with a learning rate of $10^{-5}$ using batches of 32 samples. Models are trained up to 100 epochs, but early stopping is applied so that training is halted if the validation MSE has not decreased for 10 epochs. Early stopping is used not only as a regularization techinque \citep{nakkiran2019sgd} but also to reduce the computational cost of training several thousand neural networks.

After the model for predicting the point estimates for forecasted sea level is trained, we train a separate network using the residuals $\boldsymbol{y} - \boldsymbol{\hat{\mu}}$ as targets to quantify the uncertainty associated with the prediction \citep{nix1994estimating}. Given an input $x_i$ and model parameters $\boldsymbol{\theta}$, the uncertainty network outputs a predicted uncertainty as a log-variance, $\log{\hat{\sigma_i}^2} = \log\left(\hat{\sigma}(x_i|\boldsymbol{\theta})^2\right)$. (The network outputs uncertainties as a log-variance to enforce the nonnegativity of the predicted standard deviation.) In this network, the output uncertainty is optimized using the Gaussian negative log-likelihood loss function instead of the MSE. The Gaussian likelihood of a set of independent, identically distributed observations $\boldsymbol{y} = (y_1, \dots, y_N)$ given parameters $\boldsymbol{\hat{\mu}} = (\hat{\mu}_1, \dots, \hat{\mu}_N)$ and $\boldsymbol{\hat{\sigma}}^2 = (\hat{\sigma}_1^2, \dots \hat{\sigma}_N^2)$ is given by
\begin{equation}
    p(\boldsymbol{y}| \boldsymbol{\hat{\mu}}, \boldsymbol{\hat{\sigma}}^2) = \prod_{i=1}^N \frac{1}{\hat{\sigma}_i\sqrt{2\pi}} \exp \left[-\frac{1}{2} \left(\frac{y_i - \hat{\mu}_i}{\hat{\sigma}_i}^2 \right)\right]
\end{equation}
The negative log-likelihood over all samples is given by
\begin{align}
    \begin{aligned}
    \mathcal{L}_{\textrm{NLL}}(\boldsymbol{y}| \boldsymbol{\hat{\mu}}, \boldsymbol{\hat{\sigma}}^2) 
    &= - \log \left(p(\boldsymbol{y}| \boldsymbol{\hat{\mu}}, \boldsymbol{\hat{\sigma}}^2)\right) \\ 
    &=  \frac{1}{2}\sum_{i=1}^N \left[ \log(\hat{\sigma}_i^2) + \left(\frac{y_i - \hat{\mu}_i}{\hat{\sigma}_i}\right)^2\right] + C
    \end{aligned}
    \label{eq:gnll_loss}
\end{align}
where $C = \frac{N}{2} \log(2\pi)$ is an immaterial constant under optimization. During training, the model parameters are adjusted to optimize the predicted standard deviations $\boldsymbol{\hat{\sigma}}$ given the residuals $\boldsymbol{y} - \boldsymbol{\hat{\mu}}$. 

The residuals $\boldsymbol{y} - \boldsymbol{\hat{\mu}}$ are further standardized prior to training the uncertainty network. The network architecture, optimizers, regularization, and training procedure are identical to that of the mean network, except that a learning rate of $10^{-6}$ is used for the uncertainty network to ensure convergence.

\begin{figure*}[!ht]
    \includegraphics[width=\textwidth]{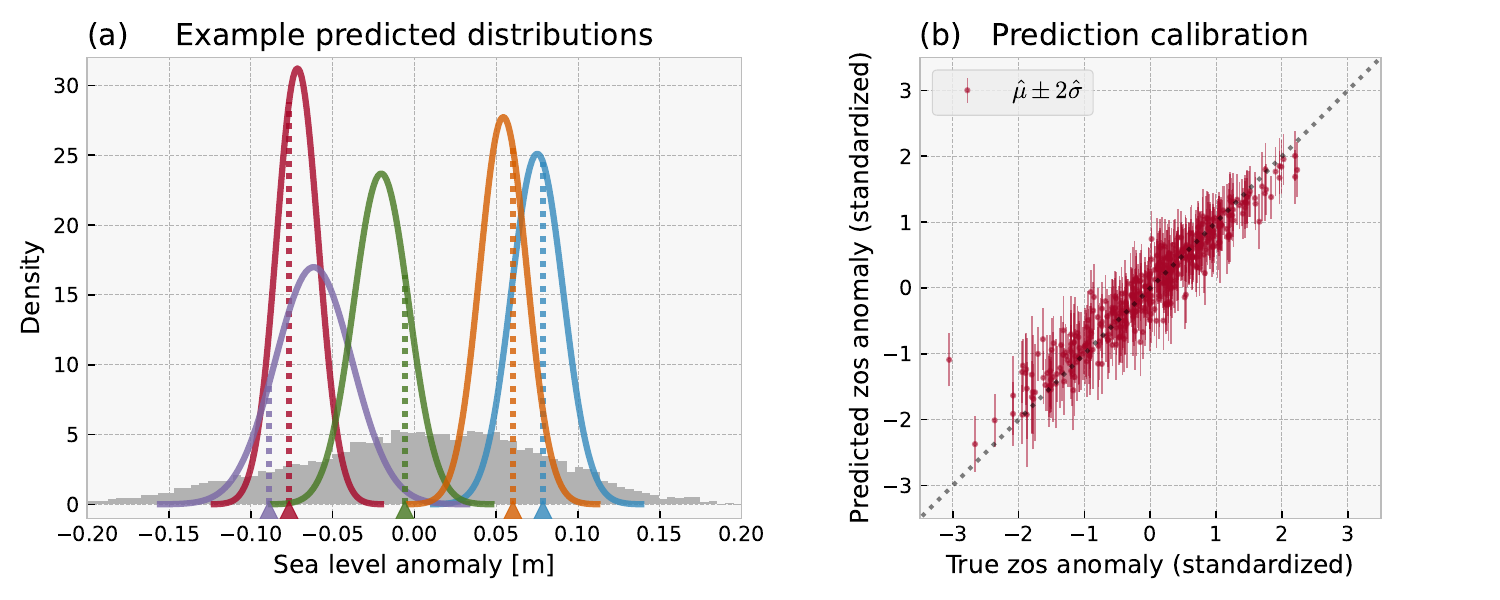}
    \caption{Example uncertainty-quantifying predictions given by a network trained at Guam (\pointA) for forecast lead $\tau=20$ days. Panel (a) shows five example distributions predicted by the machine learning framework on the test set (solid lines), compared to the climatological distribution (gray histogram). The markers on the x-axis and vertical dotted lines indicate the true target sea level anomaly. Panel (b) shows a scatterplot of 200 random dynamic sea level predictions (mean $\pm$ two standard deviations) compared to the target dynamic sea level anomaly.}
    \label{fig:uq_predictions}
\end{figure*}

Predicting a mean and standard deviation means that predictions for dynamic sea level are formulated as Gaussian probability density functions, as shown in Figure~\ref{fig:uq_predictions}a. Although the point estimate may deviate from the target value of sea level, the uncertainty-quantifying network aims to quantify the error in the forecast by a predicted standard deviation $\hat{\sigma}$. This is shown in the scatterplot in Figure 2b, where predicted dynamic sea levels $\hat{\mu}$ and predicted uncertainties $\hat{\sigma}$ are plotted against the target dynamic sea level anomaly. While the point estimate typically gives an imperfect prediction for the dynamic sea level, the uncertainty-quantifying network leverages information from the initial conditions to help quantify the forecast uncertainty.

In general, the predicted standard deviation cannot directly predict the error $y_i - \hat{\mu}_i$ for a particular sample, since the uncertainty-quantifying and point-estimate networks use identical inputs. However, it is expected that well-calibrated forecasts will have predicted standard deviations that statistically quantify the residuals when aggregating over many samples. For instance, for the particular network trained in Figure~\ref{fig:uq_predictions}b, the forecasted dynamic sea level $\hat{\mu}_i$ is within one standard deviation $\hat{\sigma}_i$ of the target value 62.7\% of the time, and within $2\hat{\sigma}_i$ 90.8\% of the time when evaluated over the test dataset. These percentages closely match the probabilities given by a normal distribution (which contains 68.3\% of its probability mass within one standard deviation of the mean and 95.4\% within two standard deviations), indicating the forecasts are well-calibrated. Thus, the forecasted standard deviation gives an estimate of the acceptable level of forecast error. For this reason, we refer to predictions with a lower $\hat{\sigma}$ as ``lower-uncertainty” or ``higher confidence" predictions, although it will in general need to be verified that such low-uncertainty predictions result in lower errors on average.

Because the dynamic sea level forecasts are given as Gaussian distributions, these forecasts can be used to evaluate probabilities of specific events. For a given probabilistic event $A$, the implied probability by the neural network framework is 
\begin{equation}
    \hat{P}_i(A) = \hat{P}_i(Y \in A) = \int_A f(y|\hat{\mu}_i, \hat{\sigma}^2_i) \ dy
\end{equation}
where $Y$ is the random variable representing the target dynamic sea level and $f(y|\hat{\mu}_i, \hat{\sigma}^2_i)$ is the probability density function of a normal distribution with parameters $\hat{\mu}_i$ and $\hat{\sigma}^2_i$. One class of events that is useful to examine to identify sources of predictability is exceedance events. For instance, the implied probability of a positive sea level anomaly is given by
\begin{equation}
    \hat{P}_i(Y \geq 0) = \int_0^\infty f(y|\hat{\mu}_i, \hat{\sigma}^2_i) \ dy.
    \label{eq:predicted_prob_pos}
\end{equation}
Nevertheless, whether the forecasted Gaussians can be used to skillfully predict probabilistic events must also be checked.

Neural networks trained on maximum-likelihood based loss functions were first proposed by \citet{nix1994estimating} and have been previously used several times in climate science to estimate stochastic uncertainties \citep{guillaumin2021stochastic, barnes2021controlled, gordon2022incorporating, barnes2023sinh}.  Typically, an assumption is made about the underlying sampling distribution for the uncertainties, and networks are trained to predict all parameters of the probability distribution together. In practice, training times were much shorter when training a network for the mean and standard deviation separately---an important practical consideration when training networks at several thousand locations to build spatial coverage. This two-stage training procedure is similar to the approach in \citet{adler2018deep} and \citet{perezhogin2023generative}. We also note that using the MSE as a training loss for the point-estimate network, as opposed to other popular regression loss functions such as the Mean Absolute Error (MAE) or Huber loss \citep{huber1964robust}, is appropriate from a theoretical perspective, as the maximum-likelihood estimator for the mean under independent, identically distributed Gaussian samples is also the argument minimizer of the MSE. Section~\ref{sec:results}\ref{subsec:network_performance} explores the validity of the two-stage training procedure by assessing the networks' probabilistic predictions. 

\subsection{Baselines: damped persistence and simple logistic regression}
\label{subsec:baseline}
As a baseline, we compare our networks to forecasts produced by damped persistence \citep[DP, ][]{lorenz1973existence, feng2025indications, long2021seasonal}.  Given local observations of dynamic sea level at a given time $x(t) \in \mathbb{R}$, the damped persistence forecast at lead $\tau$, $\hat{x}(t+\tau) \in \mathbb{R}$, is given by
\begin{equation}
    \hat{x}(t + \tau) = \beta_\tau x(t),
\end{equation}
where the autocorrelation coefficient $\beta_\tau \in [0, 1]$ is the best intermediate between a climatological forecast ($\beta_\tau = 0$) and a pure persistence forecast ($\beta_\tau = 1$). Thus, the damped persistence forecast encompasses the local linear predictability of sea level. Note that the features used in the damped persistence model have the same resolution as the target, so while the neural networks use input features at $5^\circ$ resolution to predict sea level at $1^\circ$, the damped persistence model uses $1^\circ$ inputs and outputs. The autocorrelation coefficient is computed via least squares estimation to minimize $\|x(t_i+\tau) -\beta_\tau x(t_i)\|_2$ over all times $t_i$. Autocorrelation coefficients are computed for each ensemble member in the training set and then averaged over all members.

The damped persistence model is simple and is commonly used for demonstrating deterministic forecast skill. However, because our uncertainty-quantifying framework is probabilistic, we also consider a probabilistic baseline to assess probabilistic forecasts. We employ a logistic regression baseline which uses the same features as the damped persistence model. Given an event $A$, the forecasted probability of the event occurring $\hat{P}_A(t + \tau)$ is given by
\begin{equation}
    \hat{P}_A(t + \tau) = \frac{1}{1 + e^{-\left(\beta^{(0)}_\tau + \beta^{(1)}_\tau x(t) \right)}} 
\end{equation}
where $\beta^{(0)}_\tau$ and $\beta^{(1)}_\tau$ are coefficients also determined by least-squares. The logistic regression baseline, like the damped persistence model, characterizes local predictability of dynamic sea level, but for probabilistic events. The coefficients of the logistic regression model are fitted using the Scikit-learn package \citep{pedregosa2011sklearn} with the default $l_2$ regularization weight 1.

\subsection{Metrics}
\label{subsec:metrics}
\subsubsection{Deterministic metrics}
To evaluate the skill of the point-estimate networks, we consider the Mean Absolute Error (MAE) between observations $\boldsymbol{y} = (y_1, \dots, y_N)$ and predictions $\boldsymbol{\hat{y}} = (\hat{y}_1, \dots, \hat{y}_N)$:
\begin{equation}
    \mathrm{MAE}(\boldsymbol{y}, \boldsymbol{\hat{y}}) = \frac{1}{N} \sum_{i=1}^N |y_i - \hat{y}_i|.
\end{equation}
In order to control for differences in sea level variability between different locations, the MAE is computed over standardized predictions and observations to quantify the portion of the variance the ANN is able to predict.

To directly compare the point-estimate neural networks against climatological and damped persistence baselines, we will use Mean Square Error skill scores \citep[SS, ][]{murphy1988skill}:
\begin{equation}
    \mathrm{SS} = 1 - \frac{\mathrm{MSE}_\mathrm{ANN}}{\mathrm{MSE}_\mathrm{baseline}}
\end{equation}
where $\mathrm{MSE}_f$ gives the mean square error (Eq.~\ref{eq:mse_loss}) of model $f$. The SS can be interpreted as the percentage improvement in skill gained by using the ANN as a model relative to the baseline. If the skill score is 1, the ANN has perfect skill; if the skill is 0 (or lower) the ANN has no advantage over the baseline (or is worse). Notably, the skill score relative to climatology is the same as the the neural network's coefficient of determination.

\subsubsection{Probabilistic metrics}
To complement the deterministic evaluations of the point-estimate network, we  also evaluate the networks' probabilistic prediction performance. To assess the probabilistic forecasts, we compute the Continuous Ranked Probability Score (CRPS) of the forecasts averaged over all of the samples \citep{matheson1976scoring, gneiting2007strictly, brocker2012evaluating}. Given a set of predicted cumulative distribution functions $\boldsymbol{\hat{F}} = (\hat{F}_1, \dots, \hat{F}_N)$ and observations $\boldsymbol{y} = (y_1, \dots, y_N)$, the CRPS is defined by the $L^2$-distance between the predicted and empirical distributions:
\begin{equation}
    \mathrm{CRPS}(\boldsymbol{\hat{F}}, \boldsymbol{y}) = \frac{1}{N} \sum_{i=1}^N \int_{-\infty}^\infty \left(\hat{F}_i(x) - \mathbb{1}_{\{x > y_i\}} \right)^2 \ dx.
    \label{eq:crps}
\end{equation}
The predicted cumulative distribution functions given by the neural network are Gaussian with parameters determined by the neural networks $\hat{F}_i(y) = \Phi(\hat{\sigma}_i^{-1}(y - \hat{\mu}_i))$, where $\Phi$ is the cumulative distribution function of the standard normal distribution.

The CRPS is often used for assessing probabilistic forecasts against observations. It is a strictly-proper scoring rule, meaning that probabilistic forecasts cannot artificially improve CRPS by hedging for different outcomes. Thus, CRPS penalizes overconfident predictions as well as underconfident predictions. To normalize CRPS at each location, we compute a continuous ranked probability skill score (CRPSS) relative to climatology by
\begin{equation}
    \mathrm{CRPSS}(\boldsymbol{\hat{F}}, \hat{\Psi}) = 1 - \frac{\mathrm{CRPS}(\boldsymbol{\hat{F}}, \boldsymbol{y})}{\mathrm{CRPS}(\hat{\Psi}, \boldsymbol{y})},
    \label{eq:crpss}
\end{equation}
where $\hat{\Psi}$ is the probabilistic forecast given by assuming a normal distribution with the climatological mean and standard deviation. Thus, CRPSS close to 1 indicates perfect forecast skill, while CRPSS close to 0 indicates predictions no better than climatological forecasts. The CRPS is computed using the properscoring module in Python 3 \citep{barrett2015properscoring}.

We also compute the Brier score, which is useful for assessing a model's ability to predict probabilistic events with binary outcomes \citep{brier1950verification}. The Brier score of a set of Bernoulli probability forecasts $\boldsymbol{\hat{P}} = (\hat{P}_1, \dots, \hat{P}_N)$ against binary observations of class occurrences $\boldsymbol{o} = (o_1, \dots, o_N) \in \{0, 1\}$ is given by 
\begin{equation}
    \mathrm{BS}(\boldsymbol{\hat{P}}, \boldsymbol{o}) = \frac{1}{N}\sum_{i=1}^N \left(\hat{P}_i - o_i\right)^2
    \label{eq:brier}
\end{equation}

A Brier skill score, BSS, relative to a baseline model $\boldsymbol{\hat{Q}}$, may also be defined by
\begin{equation}
    \mathrm{BSS}(\boldsymbol{\hat{P}}, \boldsymbol{\hat{Q}}) = 1 - \frac{\mathrm{BS}(\boldsymbol{\hat{P}}, \boldsymbol{o})}{\mathrm{BS}(\boldsymbol{\hat{Q}}, \boldsymbol{o})}
    \label{eq:bss}
\end{equation}
Like the CRPSS, BSS near 1 implies perfect probabilistic forecasts, while BSS near 0 indicates forecasts no better than the baseline.

\section{Results}
\label{sec:results}

\subsection{Forecast performance}
\label{subsec:network_performance}

\subsubsection{Deterministic performance and state-dependent predictability}
\label{subsubsec:deterministic_performance}

\begin{figure*}[!ht]
    \centering
    \includegraphics[width=\textwidth]{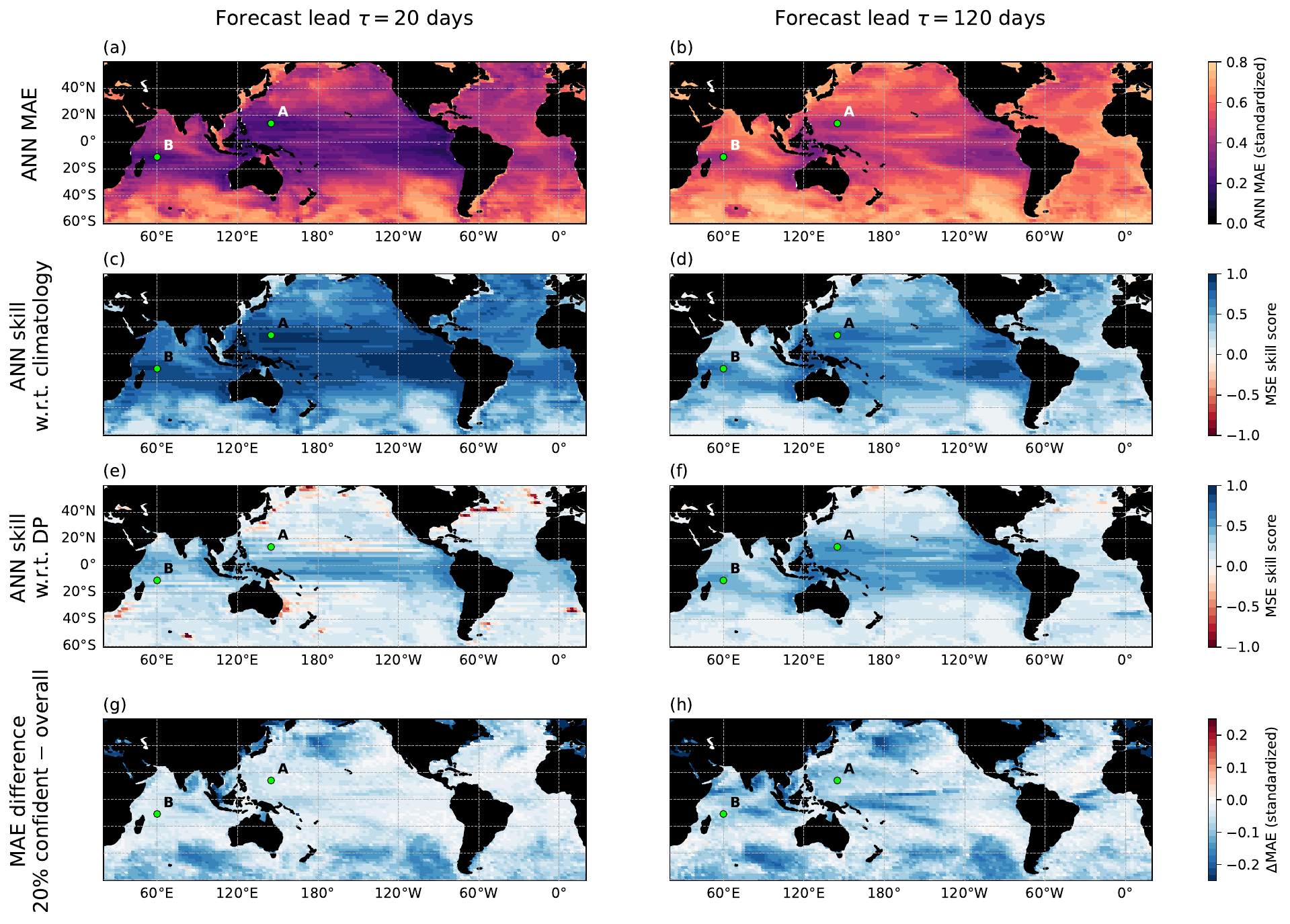}
    \caption{Global ANN prediction error metrics for forecast leads of $\tau=20$ days (a, c, e) and $\tau=120$ days (b, d, f). Note that predictions are made at every other gridpoint latitude and longitude on the nominal $1^\circ$ grid, but visualized here as a continuous map using nearest-neighbor interpolation. (a, b) Standardized mean absolute errors of the point-estimate networks over all samples. (c, d) Difference between damped persistence MAE and ANN MAE (positive values indicate ANN outperforms damped persistence. (e, f) Difference in ANN mean absolute errors taken over all samples and mean absolute errors taken over the 20\% most-confident predictions as decided by the uncertainty-quantifying network.}
    \label{fig:error_maps}
\end{figure*}

Metrics evaluating the deterministic performance of all 6,590 of the networks trained at forecast leads of $\tau=20$ and $\tau=120$ days are shown in Figure~\ref{fig:error_maps}. Figures~\ref{fig:error_maps}a and \ref{fig:error_maps}b show the MAE of the ANN's standardized predictions at each location, while Figures~\ref{fig:error_maps}c and \ref{fig:error_maps}d show the MSE skill score relative to climatology. At forecast leads of $\tau=20$ days (Fig.~\ref{fig:error_maps}c), the highest skill relative to climatology occurs primarily in the low-latitude Pacific and Indian Oceans, and the Pacific and Indian Ocean eastern boundaries bordering North America and Australia. ANN skill is low in the Southern Ocean, where ocean dynamics are dominated by baroclinic instabilities. Low forecast errors persist up to leads of $\tau=120$ days in the eastern and western tropical Pacific and tropical Indian Ocean (Fig.~\ref{fig:error_maps}d). Notably, the regions of skill in the tropical and subtropical Indo-Pacific are consistent with the regions in which dynamical models have been found to have the highest seasonal forecasting skill for observations \citep{long2021seasonal}.

To contextualize the ANN performance against the damped persistence model, Figures~\ref{fig:error_maps}c and \ref{fig:error_maps}f show the skill score of the ANN relative to damped persistence. Even by forecast leads of $\tau=20$ days, the neural networks outperform damped persistence forecasts in the majority (92.8\%) of locations (Fig.~\ref{fig:error_maps}c). Regions where the ANN does not outperform damped persistence mostly occur along western boundary currents like the Gulf Stream, Kuroshio and Agulhas Current, where spatial gradients are relatively strong and the coarse-resolution inputs of the ANN may not accurately describe the local dynamic sea level. Nevertheless, since damped persistence forecasts represent the local, linear predictability of sea level, the skill of the network relative to damped persistence indicates that nonlocal and/or nonlinear dynamics impact dynamic sea level on daily-to-seasonal timescales in most locations. The most prominent regions of high ANN skill occur adjacent to the equator, possibly due to equatorial Rossby or Kelvin waves, which propagate dynamic sea level anomalies parallel to the equator. As forecast leads increase to $\tau=120$ days, the regions of high skill expand towards higher latitudes throughout the tropics. This could reflect the fact that the phase velocity of Rossby waves decreases with latitude \citep{salmon1998lectures, rossby1939relation}, so that longer lead times are needed to transmit nonlocal sea level signals to the forecast location. The proportion of locations where the ANN outperforms damped persistence also increases to 98.6\%.

As mentioned previously, the purpose of the uncertainty-quantifying network is to identify initial conditions which may result in better forecasts. However, whether the networks are truly able to identify such initial conditions must be verified. Therefore, Figures~\ref{fig:error_maps}g and \ref{fig:error_maps}h show the difference between ANN MAE computed over all predictions and the MAE of the predictions of only the 20\% most-confident predictions from the test dataset. In 99.4\% of locations at leads of $\tau=20$ days and 98.2\% of locations at leads of $\tau=120$ days, the MAE for the 20\% of predictions deemed the most-confident by the uncertainty-quantifying network is lower than the MAE over all predictions. Thus, the networks are able to identify state-dependent predictability at nearly all locations on these timescales. Notably, many of the regions of the largest state-dependent predictability identified by the uncertainty-quantifying network occur in complementary regions to the ANN average skill relative to damped persistence, such as in the midlatitude Pacific and Southern Ocean. Therefore, the most confident predictions made by the neural network framework are often significantly better than the average damped persistence prediction.

\begin{figure*}[ht!]
    \centering
    \includegraphics[width=\textwidth]{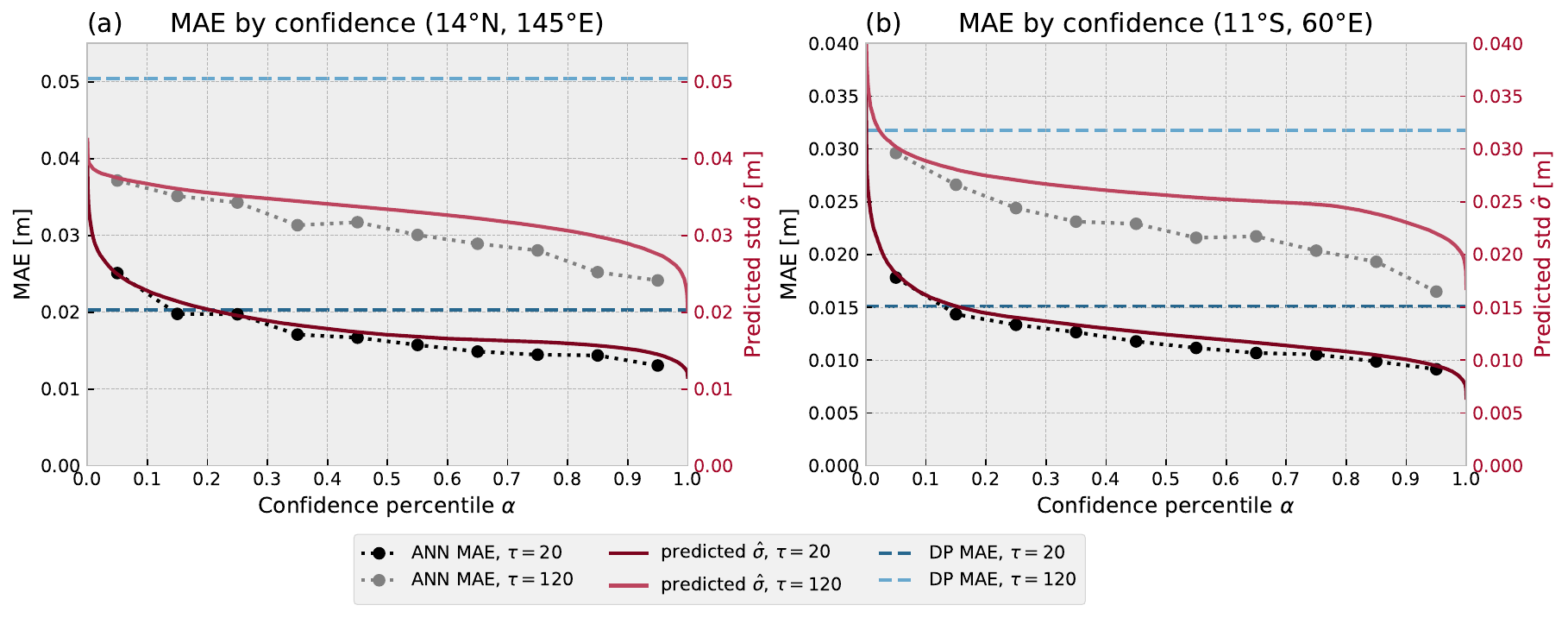}
    \caption{Test dataset mean absolute error (in meters) by network confidence level for networks trained at Guam (\pointA, panel a) and in the western Indian Ocean (\pointB, panel b). The red curve shows the predicted standard deviation outputted by the uncertainty-quantifying network as a function of the network confidence level (which is defined implicitly by the percentiles of predicted standard deviations in the test set). The black circles show the network MAEs for predictions grouped by network confidence decile. The blue line shows the MAE for the damped-persistence model. Dark colors indicate errors and uncertainties for forecast leads of $\tau=20$ days, whereas lighter shades indicate forecast leads of $\tau=120$ days.}
    \label{fig:mae_by_confidence}
\end{figure*}

Figure \ref{fig:mae_by_confidence} clarifies the relationship between predicted uncertainty and prediction errors for networks trained at Guam (\pointA) and in the western Indian Ocean (\pointB), indicated by the green dots labelled ``A" and ``B," respectively, in Figure \ref{fig:error_maps}. While the correspondence is not exact, the predicted uncertainty approximates the prediction MAE for each confidence decile. Figure~\ref{fig:mae_by_confidence} also demonstrates the utility of focusing on state-dependent predictability. For instance, although the average point-estimate neural network prediction is marginally better than damped persistence at forecast leads of $\tau=20$ days at both locations, the most-confident predictions are significantly better. As forecast leads increase to $\tau=120$ days, the difference between the neural network framework and damped persistence is even more apparent. Not only does the average network prediction outperform the baseline, but also the difference between the errors for the most-confident predictions and the median-confidence predictions increases for both of these locations.

\subsubsection{Probabilistic predictions}
\label{subsec:results:probabilistic_predictions}

\begin{figure*}[!ht]
    \centering
    \includegraphics[width=\linewidth]{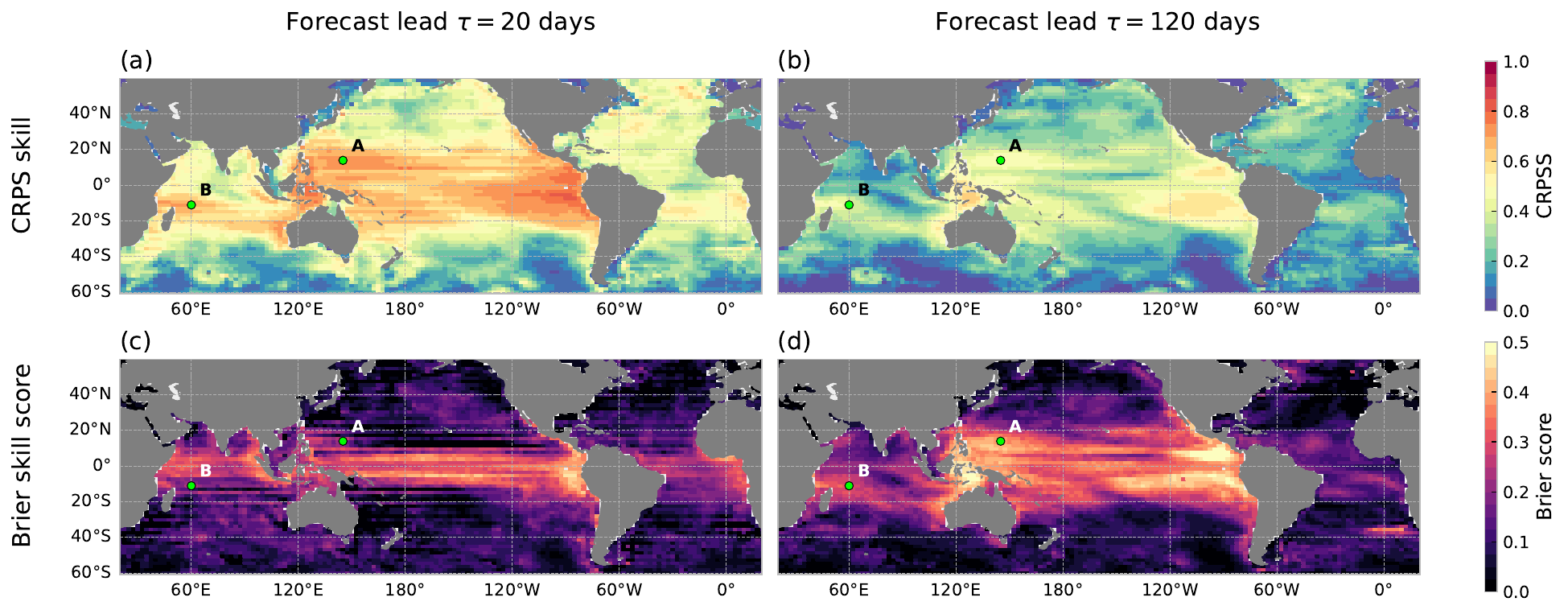}
    \caption{Global probabilistic performance metrics for networks trained at forecast leads of $\tau = 20$ days (a, c) and $\tau = 120$ days (b, d). (a, b) Continuous ranked probability skill score (Eq.~\ref{eq:crpss}) of the neural network framework evaluated relative to climatology. (c, d) Brier skill scores (Eq.~\ref{eq:bss}) for predictions of positive anomalies made by the networks evaluated relative to the logistic regression baseline discussed in Section~\ref{sec:methods}\ref{subsec:baseline}.}
    \label{fig:crps_brier}
\end{figure*}

CRPSS for all networks trained at forecast leads of $\tau = 20$ days and $\tau = 120$ days are shown in Figure \ref{fig:crps_brier}a and \ref{fig:crps_brier}b, respectively. Regions of high CRPSS occur mostly in the low-latitude Pacific and Indian Ocean. The spatial distribution of CRPSS is very highly correlated to the maps of MAE in Figure~\ref{fig:error_maps}a and \ref{fig:error_maps}b ($R^2=0.988$ at leads $\tau=20$ days and $R^2=0.963$ at leads $\tau=120$).  Such high correlations are expected, due to the fact that the MAE is a special case of the CRPS for deterministic predictions, where $\hat{F}_i(y)$ is represented by Heaviside functions $H(y - \hat{y}_i)$. 

Figure~\ref{fig:crps_brier}c and \ref{fig:crps_brier}d shows the Brier skill scores of the neural networks' ability to predict positive anomalies (as defined by Equation~\ref{eq:predicted_prob_pos}) relative to the logistic regression baseline. Brier skill scores are greater than 0 at most locations (65.8\% of locations at leads of $\tau = 20$ days and $70.0\%$ of locations at $\tau=120$ days). That is, although the neural network framework has not been explicitly tasked with predicting positive sea level anomaly events, probabilities implied using the networks' predicted mean and standard deviation are better at predicting positive sea level anomalies than a logistic regression baseline which has been explicitly trained for this purpose at most locations. Of course, the neural network uses spatial information which is not available to the logistic regression model. Nevertheless, the skill of the neural networks for predicting positive anomalies illustrates how the uncertainty-quantifying neural network framework can be used for predicting exceedance events. Regions of high BSS expand from the low-latitude Indo-Pacific at forecast leads of $\tau = 20$ days towards higher latitudes at $\tau=120$. This indicates regions in which nonlocal factors become important for forecasting exceedance probabilities.

Due to the high CRPSS and BSS and low MAE at the locations labeled ``A" and ``B" in Figure~\ref{fig:error_maps} and \ref{fig:crps_brier} (\pointA\ and \pointB, respectively), the remainder of this paper focuses on these locations. Particular attention is given to identifying how drivers of sea level predictability change over different daily-to-seasonal forecast leads and what these primary drivers are.


\subsection{Predictability by forecast lead}
\label{subsec:foo}

\begin{figure*}
    \centering
    \includegraphics[width=\textwidth]{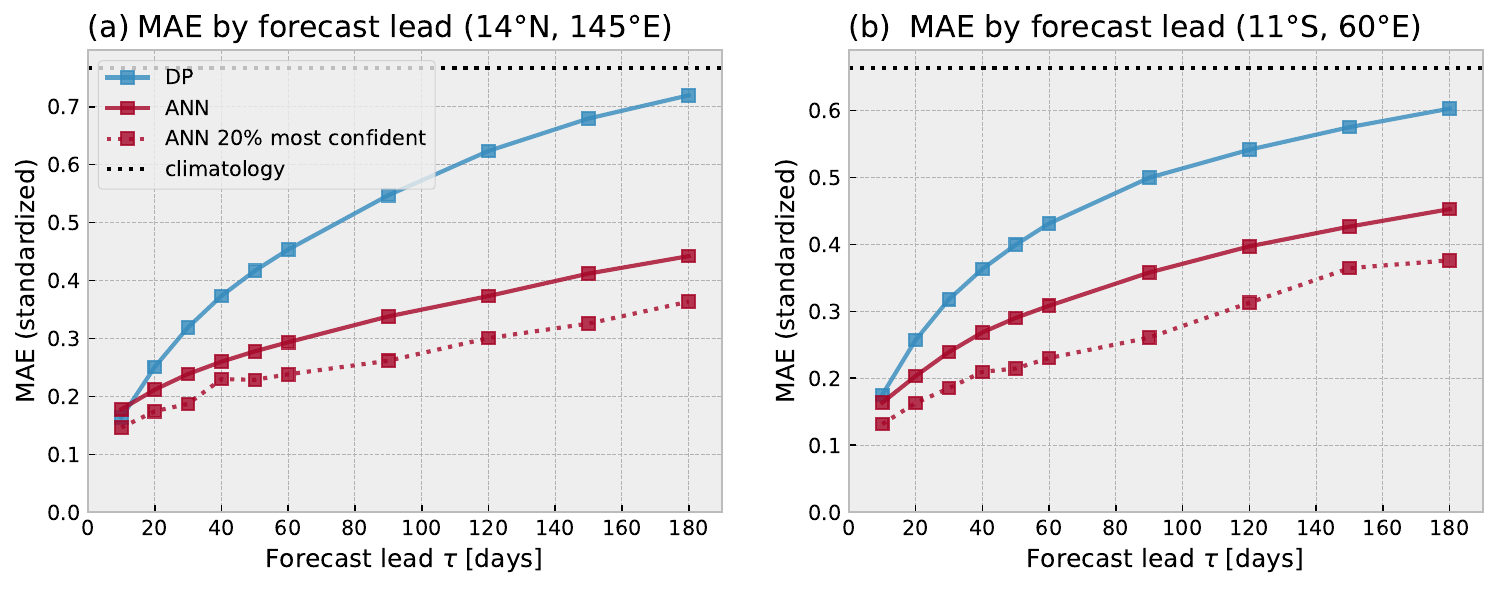}
    \caption{Mean absolute error by forecast lead time for networks trained at Guam (\pointA, panel a) and in the western Indian Ocean, (\pointB, panel b). Markers indicate forecasting leads at which networks have been trained. Black dashed line shows the climatological error, while the blue curve shows the damped persistence error as a function of lead time. The red solid line shows the Mean Absolute Error of the ANN over all samples, and the red dashed line shows the errors of only the 20\% most confident predictions.}
    \label{fig:errors_by_lag}
\end{figure*}

Figure~\ref{fig:errors_by_lag} shows the performance of each of the deterministic forecasting techniques on a variety of lead times from $\tau=10$ to $\tau=180$ days at Guam (location ``A", \pointA) and the western Indian Ocean (location ``B", \pointB). Over daily-to-seasonal timescales, damped persistence errors decay towards climatological errors at both locations. Although the predictions for the neural network have similar errors to damped persistence at forecast leads of 10 or 20 days, ANN errors grow more slowly. ANN skill peaks at about 120 days at Guam with 40.2\% improvement in MAE over damped persistence, while it peaks at 28.5\% skill at 60 days in the western Indian Ocean. The 20\% most-confident predictions also have lower MAE than the average predictions at all time lags. This shows that focusing on state-dependent predictability can consistently offer advantages for forecasts at these locations.

\begin{figure*}
    \centering
    \includegraphics[width=\textwidth]{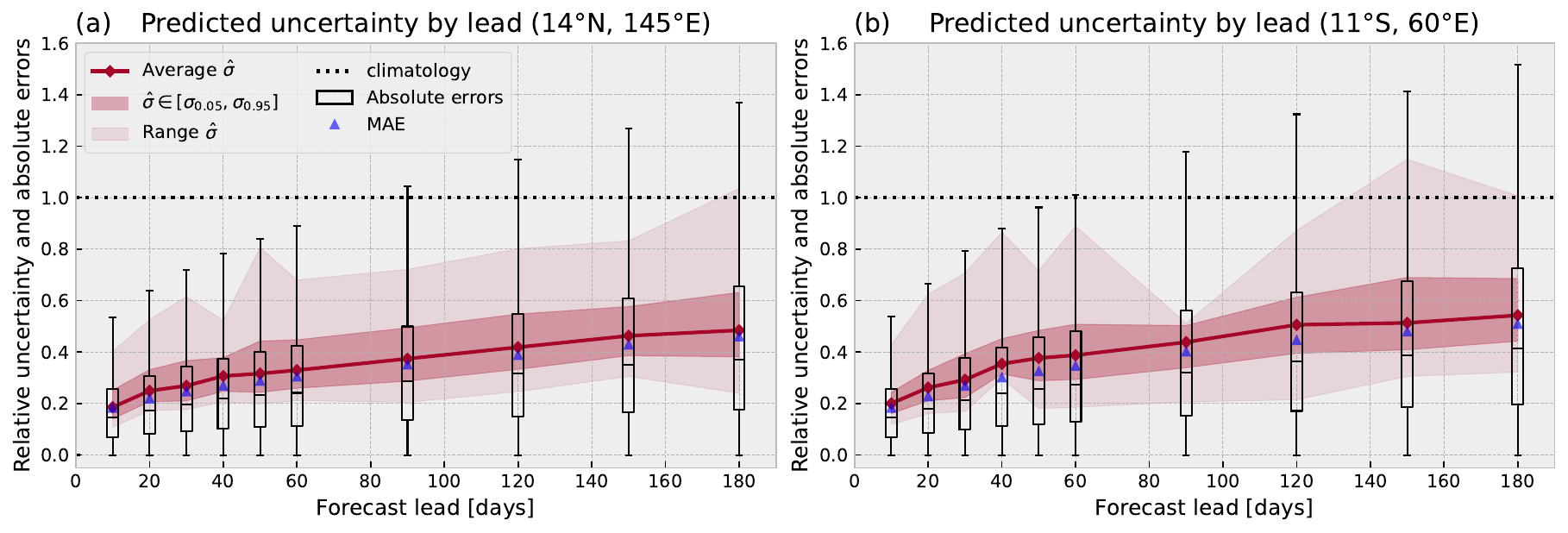}
    \caption{Distribution of predicted standard deviations $\hat{\sigma}_i$ as a function of forecast lead at Guam (\pointA, panel a) and in the Western Indian Ocean (\pointB, panel b). Predicted uncertainties are normalized relative to the climatological standard deviation. Solid red line indicates the average predicted uncertainty over all samples in the test set for each forecasting lead. Dark red shading indicates the range of standard deviations between the 5th and 95th percentiles, whereas light red shading indicates the entire min-max range of predicted standard deviations. The boxplots show the distribution of the absolute errors between the true and forecasted sea level using the point-estimate network with mean absolute errors indicated using blue triangles (outliers are removed for visualization purposes).}
    \label{fig:std_by_lag}
\end{figure*}

The MAE of forecasts in Figure~\ref{fig:errors_by_lag} increase with forecast lead because of the degradation of initial condition information from boundary forcing and chaotic dynamics. Thus, the average predicted uncertainties can be expected to also increase with larger forecast leads, plateauing at the climatological uncertainty when initial condition information has fully deteriorated. Figure~\ref{fig:std_by_lag} shows the distribution of predicted uncertainties by the neural network as a function of the forecast lead. The average predicted standard deviation does indeed increase with forecast lead, closely matching the mean absolute errors of the point-estimate network predictions and remaining less than the climatological standard deviation at all forecasting leads.  Moreover, not only do the \textit{mean} absolute errors increase with forecast leads but also the \textit{spread} of the distribution of absolute errors increases with forecast leads. Accordingly, the range of predicted standard deviations increases with forecast lead. Using a $t$-distributed Wald test for positive slope finds that the increasing minimum-maximum and 90\%-interpercentile ranges of predicted standard deviations are significant at the 5\% significance level for both locations.

\begin{figure*}[ht!]
    \centering
    \includegraphics[width=\textwidth]{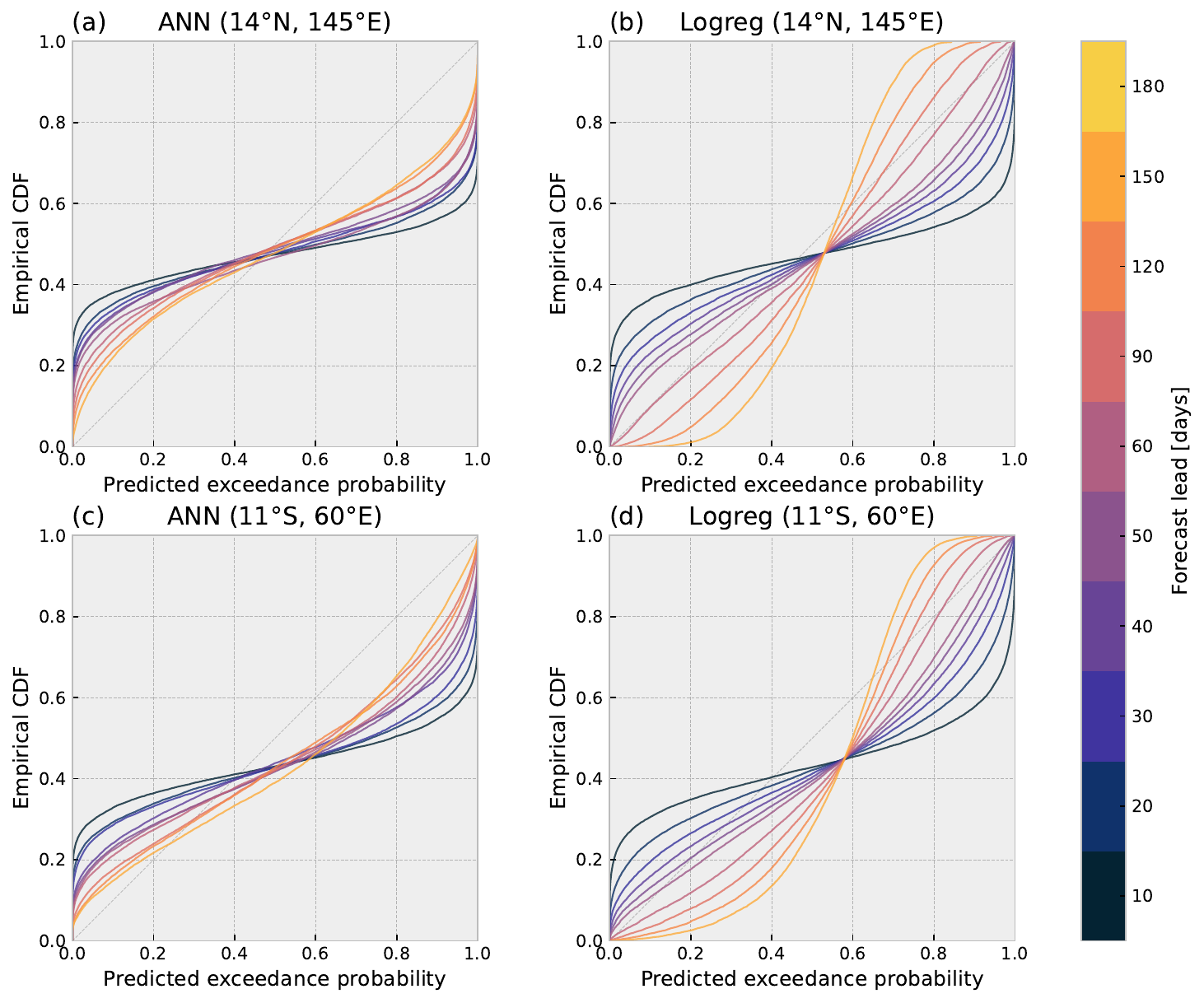}
    \caption{Empirical distribution of predicted probabilities of positive sea level anomaly exceedance events for neural network framework (a, c) and logistic regression baseline (b, d) at Guam, \pointA\ (a, b) and in the Western Indian Ocean \pointB\ (c, d).}
    \label{fig:prob_by_lag}
\end{figure*}

The increasing MAE and broadening range of errors due to the deterioration of initial condition information also have implications for the predicted probabilities of exceedance events.  For instance, predictions with completely certain exceedance outcomes would be represented by exceedance probabilities of exactly 0 or exactly 1.  On the other hand, forecasts of exceedance probabilities in which initial conditions provide no information about the outcome would simply output climatological probabilities of the event. The amount of information contained in the initial conditions for each forecast is conveyed by the empirical distributions of predicted probabilities for positive anomalies in Figure~\ref{fig:prob_by_lag}. At shorter forecast leads of $\tau=20$ days, most of the neural network predicted probabilities of positive sea level anomalies are near 0 or 1: 68\% of ANN predicted probabilities $\hat{P}_i$ yield $\hat{P}_i < 0.05$ or $\hat{P}_i > 0.95$ for location A and 63\% of probabilities are this extreme for location B. However, as forecast leads increase, the proportion of high-confidence predictions of exceedance probabilities decreases: at leads of $\tau=120$ days, this proportion of extreme probabilities has fallen to 47\% for location A and 30\% for location B. The loss of information contained in the initial conditions for the logistic regression model is much more pronounced. For instance, while the number of highly certain predictions at location A yielding probabilities of less than 0.05 or greater than 0.95 of the logistic regression model is similar to the ANN at leads of $\tau=20$ days (56\% for logistic regression vs 67\% for the ANN), by forecast leads of $\tau=120$ days, only 1\% percent of predictions are this confident for the logistic regression model whereas 47\% percent of the ANN predictions are this confident. The fact that initial condition information decays more quickly for the logistic regression baseline than for the neural networks suggests that non-local information becomes more important for predicting exceedance probabilities while local information becomes less important over daily-to-seasonal timescales.

\subsection{Drivers of predictability over different forecast leads}
\label{subsec:drivers}

\begin{figure*}[ht!]
    \centering
    \includegraphics[width=\textwidth]{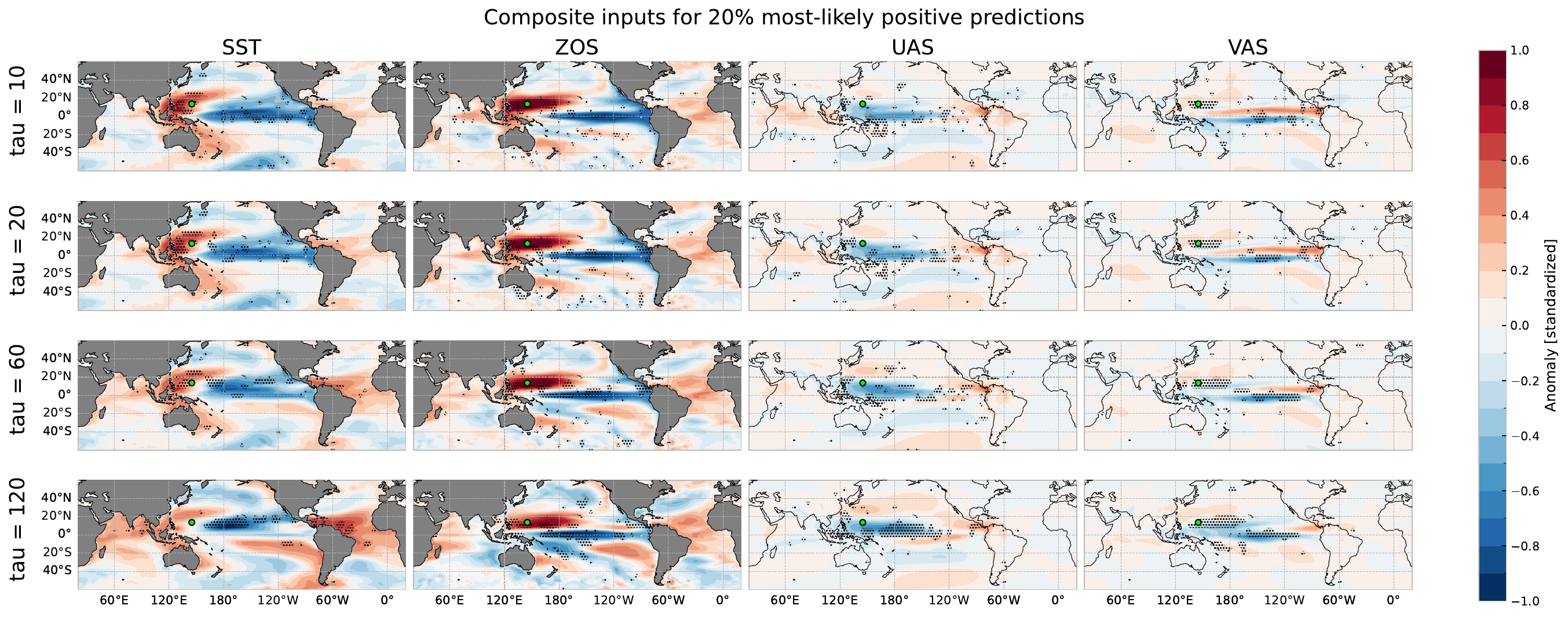}
    \caption{Composites averaged over the 20\% of input samples most likely to result in positive sea level anomalies at Guam (\pointA, green dot) for different forecast leads. Columns show different input fields (SST, ZOS, UAS, VAS), and rows show different forecast leads ($\tau=10$, 20, 60, and 120 days). Stippling indicates integrated gradients of the point-estimate network exceeding the $95^\textrm{th}$ percentile for each variable.}
    \label{fig:drivers_A}
\end{figure*}

\begin{figure*}[ht!]
    \centering
    \includegraphics[width=\textwidth]{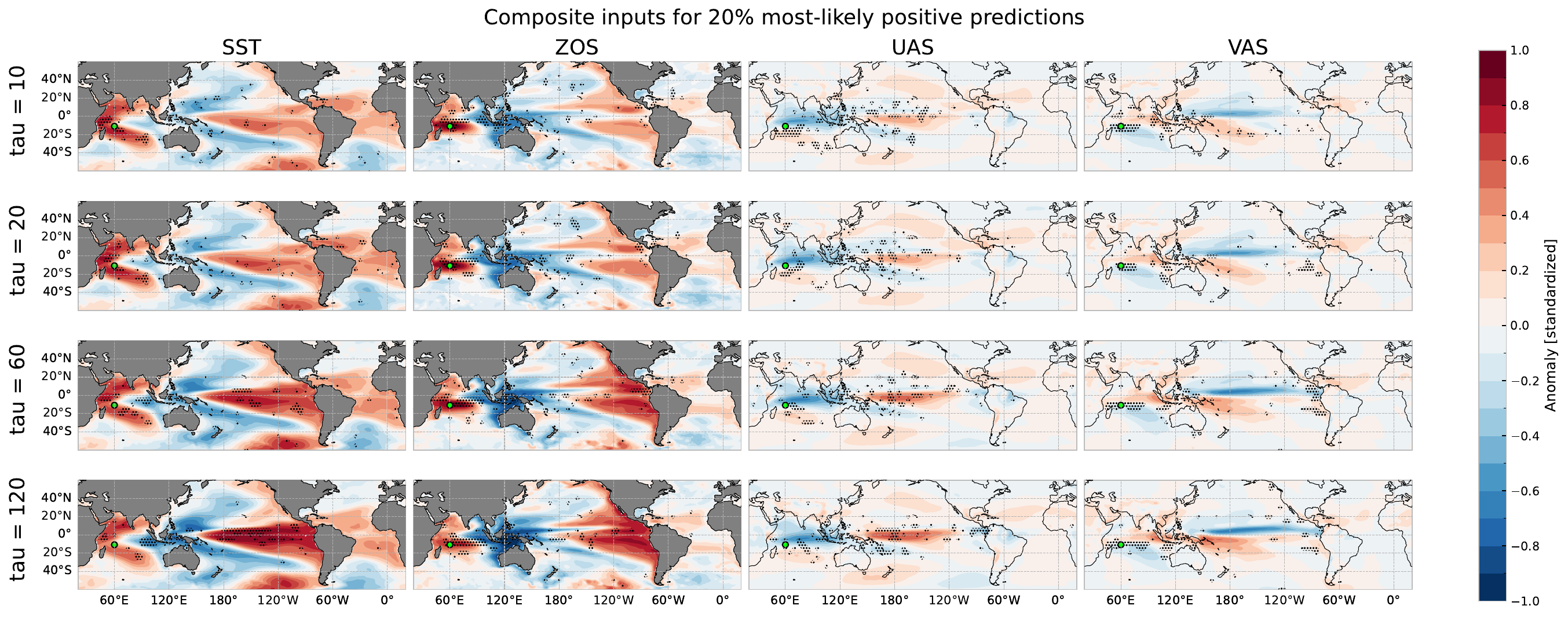}
    \caption{Same as Figure~\ref{fig:drivers_A}, but for predictions in the western Indian Ocean (\pointB).}
    \label{fig:drivers_B}
\end{figure*}

Figure~\ref{fig:prob_by_lag} suggests that both local and nonlocal drivers can impact sea level predictability on daily-to-seasonal timescales. To identify these drivers, Figures~\ref{fig:drivers_A} and ~\ref{fig:drivers_B} show the composite inputs averaged over the 20\% of samples most likely to result in positive sea level anomalies as predicted by the neural networks at Guam (\pointA) and the western Indian Ocean (\pointB), respectively. We highlight forecasting leads of $\tau=10$, 20, 60, and 120 days, loosely informed by the rates of increasing forecast error in Figure~\ref{fig:errors_by_lag}. (While thresholding on exceedance probability is also possible, it is avoided as it yields an inconsistent number of samples for different time lags.) The composites thus show the average initial condition resulting in a likely positive anomaly (using the definition of exceedance probabilities from Equation~\ref{eq:predicted_prob_pos}) . While the composites show \textit{which} samples result in likely positive anomalies, it is unclear \textit{why} the networks have selected these samples. Therefore, a neural network attribution technique, integrated gradients \citep{sundararajan2017axiomatic, mamalakis2022neural}, was applied to the point-estimate networks to quantify the relative contribution of each input feature to positive sea level anomalies. Integrated gradients are applied to the point-estimate network to identify input features that result in more positive anomalies. The $95^{\textrm{th}}$-percentile integrated gradients for each variable and forecast lead are stippled in Figures \ref{fig:drivers_A} and \ref{fig:drivers_B}.

At Guam (\pointA, Fig.~\ref{fig:drivers_A}), the most prominent initial conditions resulting in likely positive anomalies is the persistence of local sea level anomalies at all time lags. The local dynamic sea level pattern resulting in likely positive predictions is stretched significantly zonally, extending especially eastward from the prediction location. Moreover, the regional maximum-composite input dynamic sea level moves eastward with increasing forecast lead. This could be due to incident Rossby waves, which propagate sea level signals with a westward phase velocity and could be a nonlocal source of predictability \citep{vallis2017atmospheric}.

The SST fields resulting in likely-positive sea level anomalies at Guam exhibit a pattern that persists over forecast leads from 10--120 days, with positive SST anomalies northwest of the prediction location but with negative and intensifying SSTs eastward in the Central Pacific. This SST pattern is somewhat distinct from some major drivers of climate variability occurring in the Pacific, lacking the signature east Pacific tongue of the El Ni\~no-Southern Oscillation or the strong footprint of SSTs in the central North Pacific of the Pacific Decadal Oscillation. However, the SST pattern is quite similar to the region of strong seasonal lagged-correlation between Pacific SSTs and sea level recorded by tide gauges at Guam found in \citet{chowdhury2007seasonal}. Thus, larger surface ocean heat content in the western tropical Pacific drives positive sea level anomalies at Guam through dynamical mechanisms robust to both CESM2 simulations and observations.

In the western Indian Ocean (\pointB, Fig.~\ref{fig:drivers_B}),  SST and ZOS composites point to a strong signal from the Indian Ocean resembling the positive phase of the Indian Ocean Dipole \citep[IOD,][]{saji1999dipole, webster1999coupled}. The surface wind divergence east of Indonesia found in the composites is consistent with the Walker circulation that accompanies the dipole SST patterns. This indicates that the IOD is a significant source of information for the likelihood of sea level anomalies in the western Indian Ocean. Increases in SSTs in the western Indian Ocean during a positive IOD likely drive thermosteric sea level, driving persistent sea level anomalies and increasing the forecasted likelihood of positive anomalies on daily-to-seasonal timescales. The relationship is corroborated by \citet{roberts2016drivers}, which identifies a significant fingerprint of the IOD on observed thermosteric sea level in the Indian Ocean. The dipole signature in SST and ZOS also slightly diminishes over the daily-to-seasonal timescale. This is consistent with the timescale of the IOD index, which ranges from weeks to months \citep{wang2016evolution, behera2013tropical, rao2004abrupt}.

El Ni\~no also emerges as a prominent source of western Indian Ocean dynamic sea level predictability on daily-to-seasonal timescales, as evidenced by the intensifying central and eastern Pacific SST pattern. The integrated gradients indicate that while local SSTs are more important for predicting dynamic sea level at forecast leads of $\tau=10$ and $\tau=20$ days, tropical SSTs in the Ni\~no 3.4 region are more useful for forecasting likely positive sea level anomalies at $\tau=60$ and $\tau=120$ days. El Ni\~no events may impact Indian Ocean dynamic sea level through its basin-scale influence \citep{xie2002structure} as well as by its own influence on the IOD \citep{stuecker2017revisiting, yang2015seasonality, krishnamurthy2003variability}. Thus, even though the local predictability of western Indian Ocean dynamic sea level due to the IOD may decrease with forecast leads, El Ni\~no may provide a remote source of predictability for longer leads by being a precursor to the IOD.

\section{Discussion}
\label{sec:discussion}

Dynamic sea level on daily-to-seasonal timescales is driven by a variety of processes in the atmosphere and ocean. Identifying conditions where predictability is enhanced can reduce uncertainties and result in ``windows of opportunity" for better forecasts. In this study, uncertainty-quantifying regression neural networks were trained on CESM2 coupled climate data using the Gaussian maximum log-likelihood to identify sources of state-dependent predictability for dynamic sea level. The uncertainty-quantifying networks’ predicted standard deviations can be leveraged to identify initial conditions that result in better forecasts at most locations and timescales. This establishes that state-dependent sources of predictability of sea level exist and demonstrates the utility of detecting these fortuitous initial conditions.

While this study has established how neural networks can be used for identifying sources of predictability, we make no claim that the trained models in this study are optimal for forecasting applications. In addition to the myriad of different hyperparameters which could be modified to improve network performance, it is possible that using different features could result in higher skill relative to damped persistence. For instance, using higher resolution inputs could provide a richer representation of the initial conditions used for the forecast, improving the skill of the network in regions such as western boundary currents in which the spatial gradients are strong. However, using higher resolution inputs also increases the dimensionality of the input space, necessitating more training samples and computational resources for model convergence. Furthermore, chaotic and nonlinear dynamics also imposes fundamental limits to the skill that can be attained in certain regions.

The performance of the uncertainty-quantifying networks may also be improved by using different features. Although the networks discriminate different uncertainty levels based on initial states, the range of predicted standard deviations is arguably rather small. For example, 90\% of the predicted standard deviations for the network trained at Guam at lead $\tau=120$ days fall between 33\% and 55\% of the climatological standard deviation (Fig.~\ref{fig:std_by_lag}). While the distribution of absolute errors may be statistically consistent with these predicted standard deviations, it is possible that using different resolutions, input features, or architectures could result in broader ranges of predicted standard deviations. Having a larger spread of predicted standard deviations could result in better estimates of the errors and enhance the utility of focusing on state-dependent sources of predictability.

In the uncertainty-quantifying framework, it was assumed that the uncertainties could be described using a Gaussian distribution and that the point-estimate and uncertainty-quantifying networks could be trained separately instead of learning all the parameters of the probability distribution simultaneously. The relatively high CRPSS in the low-latitude Indo-Pacific suggests that these assumptions are adequate in these regions. The regions of low CRPSS occurring in the midlatitudes and the Southern Ocean do not necessarily mean that these choices are invalid. Indeed, the extremely high correlation between CRPSS and MAE imply a general lack of predictability in these regions, resulting in systematic underprediction for the point estimate network \citep{murphy1973new} and overall predicted distributions similar to climatological forecasts.

Although not explicitly trained to forecast probabilities, we noted that the uncertainty-quantifying network framework implies forecasted exceedance probabilities that often outperform local logistic regression baselines, which \textit{are} explicitly constructed to predict exceedance probabilities. This reflects not only the relative quality of the probabilistic predictions and validity of assuming normal-distributed uncertainties but also the importance of nonlocal sources of predictability, which are registered in the features of the neural networks but not represented in the inputs of the damped persistence or logistic regression baselines. 

Using a regression neural network to forecast event probabilities also illustrates a useful application of the regression neural network framework of \citet{gordon2022incorporating}.  \citet{mayer2021subseasonal} and \citet{gordon2022incorporating} used different criteria to isolate sources of state-dependent predictability: whereas \citet{gordon2022incorporating} used predicted standard deviation to define predictable initial conditions, \citet{mayer2021subseasonal} focused on discrete probabilities such as exceedance events. The different architectures of the two studies make these different notions of predictability natural, as the regression networks of \citet{gordon2022incorporating} output standard deviation while the classification networks of \citet{mayer2021subseasonal} output event probabilities. However, as shown in this study, the regression framework of \citet{gordon2022incorporating} can be used to predict event probabilities, potentially making it a more flexible approach to identifying state-dependent sources of predictability. A direct comparison of the skill of the different approaches for predicting event probabilities could clarify the strengths and limitations of the regression framework.

The distribution of predicted exceedance probabilities by the neural networks and logistic regression baseline conveys the degradation of information available from the initial conditions for predicting exceedance events. For both forecast techniques, near-certain probabilistic exceedance outcomes decrease in frequency as the forecast leads increase, indicating timescales where initial condition information is lost. However, the empirical distributions devolve to climatology more slowly for the neural networks than for the logistic regression, suggesting that regional-scale information about the predictability of positive sea level anomalies continues to persist while the local information fades. This approach may be a useful way to characterize the loss of predictability on daily-to-seasonal timescales. However, it should be stressed that the predicted probabilities depend not only on the initial conditions but also on the quality of the forecast models themselves.

Potential physical sources of dynamic sea level predictability were investigated by analyzing composites of the input samples deemed most likely by the neural networks to result in positive sea level anomalies. At Guam, propagating Rossby waves were identified as a potential source of predictability for sea level, while in the western Indian Ocean, the persistence of sea level anomalies due to the Indian Ocean Dipole provided a source of predictability, though the influence of El Ni\~no begins to emerge on seasonal timescales. 

Many of the sources of state-dependent predictability identified in the composites seem to come from low-frequency drivers, such as sea surface temperatures or dynamic sea level.  Significant sources of predictability from the surface wind fields were more difficult to identify through our approach. Multiple studies have shown how surface winds can impact dynamic sea level through, for instance, manometric changes from wind stress \citep{hermans2022effect, arcodia2024subseasonal, fukumori1998nature}, surface Ekman mass convergence \citep{kamp2024tropical, piecuch2011mechanisms}, or changes in large-scale Sverdrup balance \citep{cabanes2006contributions, roberts2016drivers}. The prominence of low-frequency drivers found using our methods could be due to the averaging over multiple samples to identify robust initial states with predictable outcomes, as distinct, high frequency inputs would be filtered out. Cluster analysis methods, such as $k$-means or DBSCAN, could be used to identify distinct drivers. However, applying such methods introduces numerous sensitivities (e.g., type of clustering algorithm, dimensionality reduction techniques, and number of components) and is beyond the scope of this work. 

A final caveat of this study is the fact that the drivers of predictability found here are learned using simulations from a coupled dynamical model simulation rather than observations. Like all climate models, CESM2 contains biases against the observational record. While CESM2 captures the dominant timescales, spectrum, precursors, and teleconnections of ENSO well \citep{capotondi2020enso}, CESM2 overestimates the amplitude of El Ni\~no by approximately 30\%. CESM2 has also been observed to have high-temperature and low-salinity biases in the tropical Pacific, causing SST anomalies in the eastern and central Pacific to develop differently from ocean reanalysis \citep{wei2021tropical}. Such biases limit the direct applicability of the forecasts developed in this study to the real world. Nevertheless, the sources of predictability identified in this study could serve as a starting point of future studies for identifying state-dependent predictability and improving forecasts. For example, the influence of the IOD or ENSO as a source of dynamic sea level predictability could be further examined through observational products. In particular, transfer learning presents an opportunity to refine the relationships learned from hundreds of years of CESM2 simulation data to the real-world altimetry observations which are limited to a few decades of samples \citep{pan2010survey, eyring2024pushing}. Transfer learning on climate model data has been successfully used to improve forecasts of El Ni\~no \citep{ham2019deep} as well as long-term climate projections \citep{barnes2025combining, immorlano2025transferring}, pointing to the potential fruitfulness of such an approach.

\acknowledgments
A.B. is supported by the VoLo Foundation. L.Z. and A.B. were supported, in part, by NOAA grant NA20OAR4310411, L.Z. also acknowledges support from NOAA NA20OAR4310397 and the KITP Program “Machine Learning and the Physics of Climate” under the National Science Foundation Grant No. NSF PHY-1748958. E.A.B. is funded, in part, by NOAA grants NA24OARX431C0022 and NA22OAR4310621. We would like to acknowledge computing support from the Casper system (\url{https://ncar.pub/casper}) provided by the NSF National Center for Atmospheric Research (NCAR) \citet{casperhpc}. We thank Emily Gordon and Marybeth Arcodia for insightful discussions on this work and the constructive comments from three anonymous reviewers. 


%
%
\datastatement
The CESM2 Large Ensemble Dataset, provided by the CESM2 Large Ensemble Community Project in partnership with the IBS Center for Climate Physics in South Korea, is available from the NCAR Climate Data Gateway at \url{https://doi.org/10.26024/kgmp-c556} \citep{cesm2le}, as referenced in \citet{rodgers2021ubiquity}.  The code used for data processing, training, analysis and visualization in this study, as well as the files for reproducing the software environment, are provided under the MIT license at \url{https://github.com/andrewbrettin/zos_predictability_aies} \citep{brettin2025code}.

%






%



\bibliographystyle{ametsocV6}
\bibliography{references}

\begin{thebibliography}{110}
\providecommand{\natexlab}[1]{#1}
\providecommand{\url}[1]{\texttt{#1}}
\renewcommand{\UrlFont}{\rmfamily}
\providecommand{\urlprefix}{URL }
\expandafter\ifx\csname urlstyle\endcsname\relax
  \providecommand{\doi}[1]{https://doi.org/\discretionary{}{}{}#1}\else
  \providecommand{\doi}{https://doi.org/\discretionary{}{}{}\begingroup
  \urlstyle{rm}\Url}\fi
\providecommand{\eprint}[2][]{\url{#2}}

\bibitem[{Adler and {\"O}ktem(2018)Adler, and {\"O}ktem}]{adler2018deep}
Adler, J., and O.~{\"O}ktem, 2018: Deep bayesian inversion. \textit{arXiv
  preprint arXiv:1811.05910}.

\bibitem[{Albers and Newman(2019)Albers, and Newman}]{albers2019priori}
Albers, J.~R., and M.~Newman, 2019: {A Priori Identification of Skillful
  Extratropical Subseasonal Forecasts}. \textit{Geophys.\ Res.\ Lett.},
  \textbf{46~(21)}, 12\,527--12\,536.

\bibitem[{Amaya et~al.(2022)Amaya, Jacox, Dias, Alexander, Karnauskas, Scott,,
  and Gehne}]{amaya2022subseasonal}
Amaya, D.~J., M.~G. Jacox, J.~Dias, M.~A. Alexander, K.~B. Karnauskas, J.~D.
  Scott, and M.~Gehne, 2022: {Subseasonal-to-Seasonal Forecast Skill in the
  California Current System and Its Connection to Coastal Kelvin Waves}.
  \textit{J.\ Geophys.\ Res.\ Oceans}, \textbf{127~(1)}, e2021JC017\,892.

\bibitem[{Aparna et~al.(2012)Aparna, McCreary, Shankar,, and
  Vinayachandran}]{aparna2012signatures}
Aparna, S., J.~McCreary, D.~Shankar, and P.~Vinayachandran, 2012: {Signatures
  of Indian Ocean Dipole and El Ni{\~n}o--Southern Oscillation events in sea
  level variations in the Bay of Bengal}. \textit{J.\ Geophys.\ Res.\ Oceans},
  \textbf{117~(C10)}.

\bibitem[{Arcodia et~al.(2024)Arcodia, Becker,, and
  Kirtman}]{arcodia2024subseasonal}
Arcodia, M.~C., E.~Becker, and B.~P. Kirtman, 2024: {Subseasonal Variability of
  US Coastal Sea Level from MJO and ENSO Teleconnection Interference}.
  \textit{Wea.\ Forecasting}, \textbf{39~(2)}, 441--458.

\bibitem[{Balmaseda et~al.(2024)Balmaseda, McAdam, Masina, Mayer, Senan,
  de~Bosiss{\'e}son,, and Gualdi}]{balmaseda2024skill}
Balmaseda, M.~A., R.~McAdam, S.~Masina, M.~Mayer, R.~Senan,
  E.~de~Bosiss{\'e}son, and S.~Gualdi, 2024: Skill assessment of seasonal
  forecasts of ocean variables. \textit{Front.\ Mar.\ Sci.}, \textbf{11},
  1380\,545.

\bibitem[{Barnes and Barnes(2021)Barnes, and Barnes}]{barnes2021controlled}
Barnes, E.~A., and R.~J. Barnes, 2021: {Controlled Abstention Neural Networks
  for Identifying Skillful Predictions for Regression Problems}. \textit{J.\
  Adv.\ Model.\ Earth\ Syst.}, \textbf{13~(12)}, e2021MS002\,575.

\bibitem[{Barnes et~al.(2023)Barnes, Barnes,, and DeMaria}]{barnes2023sinh}
Barnes, E.~A., R.~J. Barnes, and M.~DeMaria, 2023: Sinh-arcsinh-normal
  distributions to add uncertainty to neural network regression tasks:
  Applications to tropical cyclone intensity forecasts. \textit{Env.\ Data
  Sci.}, \textbf{2}, e15.

\bibitem[{Barnes et~al.(2024)Barnes, Diffenbaugh,, and
  Seneviratne}]{barnes2025combining}
Barnes, E.~A., N.~S. Diffenbaugh, and S.~I. Seneviratne, 2024: {Combining
  climate models and observations to predict the time remaining until regional
  warming thresholds are reached}. \textit{Env.\ Res.\ Lett.}, \textbf{20~(1)},
  014\,008.

\bibitem[{Barrett et~al.(2015)Barrett, Hoyer, Kleeman, O'Kane
  et~al.}]{barrett2015properscoring}
Barrett, L., S.~Hoyer, A.~Kleeman, D.~O'Kane, and Coauthors, 2015:
  properscoring. The Climate Corporation, gitHub, accessed 2024-12-12,
  \url{https://github.com/properscoring/properscoring/tree/master}.

\bibitem[{Becker et~al.(2012{\natexlab{a}})Becker, Meyssignac, Letetrel,
  Llovel, Cazenave,, and Delcroix}]{becker2012sea_a}
Becker, M., B.~Meyssignac, C.~Letetrel, W.~Llovel, A.~Cazenave, and
  T.~Delcroix, 2012{\natexlab{a}}: {Sea level variations at tropical Pacific
  islands since 1950}. \textit{Global Planet.\ Change}, \textbf{80}, 85--98.

\bibitem[{Becker et~al.(2012{\natexlab{b}})Becker, Meyssignac, Letetrel,
  Llovel, Cazenave,, and Delcroix}]{becker2012sea_b}
Becker, M., B.~Meyssignac, C.~Letetrel, W.~Llovel, A.~Cazenave, and
  T.~Delcroix, 2012{\natexlab{b}}: {Sea level variations at tropical Pacific
  islands since 1950}. \textit{Global Planet.\ Change}, \textbf{80}, 85--98.

\bibitem[{Behera et~al.(2013)Behera, Brandt,, and
  Reverdin}]{behera2013tropical}
Behera, S., P.~Brandt, and G.~Reverdin, 2013: The tropical ocean circulation
  and dynamics. \textit{Ocean Circulation and Climate}, Vol. 103, Elsevier,
  385--412.

\bibitem[{Brettin(2025)}]{brettin2025code}
Brettin, A., 2025: {Code for Brettin, Zanna, and Barnes (2025):
  daily-to-seasonal dynamic sea level predictabilty (v1.0.0)}. Zenodo,
  \doi{10.5281/zenodo.14873345}.

\bibitem[{Brier(1950)}]{brier1950verification}
Brier, G.~W., 1950: Verification of forecasts expressed in terms of
  probability. \textit{Mon.\ Wea.\ Rev.}, \textbf{78~(1)}, 1--3.

\bibitem[{Br{\"o}cker(2012)}]{brocker2012evaluating}
Br{\"o}cker, J., 2012: Evaluating raw ensembles with the continuous ranked
  probability score. \textit{Quart.\ J.\ Roy.\ Meteor.\ Soc.},
  \textbf{138~(667)}, 1611--1617.

\bibitem[{Cabanes et~al.(2006)Cabanes, Huck,, and Colin~de
  Verdi{\`e}re}]{cabanes2006contributions}
Cabanes, C., T.~Huck, and A.~Colin~de Verdi{\`e}re, 2006: {Contributions of
  Wind Forcing and Surface Heating to Interannual Sea Level Variations in the
  Atlantic Ocean}. \textit{J.\ Phys.\ Oceanogr.}, \textbf{36~(9)}, 1739--1750.

\bibitem[{Capotondi et~al.(2020)Capotondi, Deser, Phillips, Okumura,, and
  Larson}]{capotondi2020enso}
Capotondi, A., C.~Deser, A.~Phillips, Y.~Okumura, and S.~Larson, 2020: {ENSO
  and Pacific decadal variability in the Community Earth System Model version
  2}. \textit{J.\ Adv.\ Model.\ Earth\ Syst.}, \textbf{12~(12)},
  e2019MS002\,022.

\bibitem[{Chen et~al.(2023)Chen, Yang,, and Wu}]{chen2023topography}
Chen, L., J.~Yang, and L.~Wu, 2023: {Topography Effects on the Seasonal
  Variability of Ocean Bottom Pressure in the North Pacific Ocean}. \textit{J.\
  Phys.\ Oceanogr.}, \textbf{53~(3)}, 929--941.

\bibitem[{Chowdhury and Chu(2015)Chowdhury, and Chu}]{chowdhury2015sea}
Chowdhury, M.~R., and P.-S. Chu, 2015: {Sea level forecasts and early-warning
  application: Expanding cooperation in the South Pacific}. \textit{Bull.\
  Amer.\ Meteor.\ Soc.}, \textbf{96~(3)}, 381--386.

\bibitem[{Chowdhury et~al.(2007{\natexlab{a}})Chowdhury, Chu,, and
  Schroeder}]{chowdhury2007enso}
Chowdhury, M.~R., P.-S. Chu, and T.~Schroeder, 2007{\natexlab{a}}: {ENSO and
  seasonal sea-level variability--a diagnostic discussion for the US-Affiliated
  Pacific Islands}. \textit{Theor.\ Appl.\ Climatol.}, \textbf{88}, 213--224.

\bibitem[{Chowdhury et~al.(2007{\natexlab{b}})Chowdhury, Chu, Schroeder,, and
  Colasacco}]{chowdhury2007seasonal}
Chowdhury, M.~R., P.-S. Chu, T.~Schroeder, and N.~Colasacco,
  2007{\natexlab{b}}: {Seasonal sea-level forecasts by canonical correlation
  analysis---an operational scheme for the U.S.-affiliated Pacific Islands}.
  \textit{Int.\ J.\ Climatol.}, \textbf{27~(10)}, 1389--1402.

\bibitem[{Christensen et~al.(2020)Christensen, Berner,, and
  Yeager}]{christensen2020value}
Christensen, H.~M., J.~Berner, and S.~Yeager, 2020: {The Value of
  Initialization on Decadal Timescales: State-Dependent Predictability in the
  CESM Decadal Prediction Large Ensemble}. \textit{J.\ Climate},
  \textbf{33~(17)}, 7353--7370.

\bibitem[{{Computational and Information Systems Laboratory}(2023)}]{casperhpc}
{Computational and Information Systems Laboratory}, 2023: {Casper: HPE Cray EX
  System (University Community Computing)}. NSF National Center for Atmospheric
  Research, Boulder, CO, \doi{10.5065/qx9a-pg09}.

\bibitem[{Danabasoglu et~al.(2021)Danabasoglu, Deser, Rodgers,, and
  Timmermann}]{cesm2le}
Danabasoglu, G., C.~Deser, K.~Rodgers, and A.~Timmermann, 2021: {CESM2 Large
  Ensemble Dataset}. National Center for Atmospheric Research,
  \urlprefix\url{https://www.earthsystemgrid.org/dataset/ucar.cgd.cesm2le.output.html},
  \doi{https://doi.org/10.26024/kgmp-c556}.

\bibitem[{Danabasoglu et~al.(2020)}]{danabasoglu2020community}
Danabasoglu, G., and Coauthors, 2020: {The Community Earth System Model Version
  2 (CESM2)}. \textit{J.\ Adv.\ Model.\ Earth\ Syst.}, \textbf{12~(2)},
  e2019MS001\,916.

\bibitem[{DeMott et~al.(2021)DeMott, Mu{\~n}oz, Roberts, Spillman,, and
  Vitart}]{demott2021benefits}
DeMott, C., {\'A}.~Mu{\~n}oz, C.~Roberts, C.~Spillman, and F.~Vitart, 2021: The
  benefits of better ocean weather forecasting. \textit{Eos}, \textbf{102}.

\bibitem[{Doi et~al.(2020)Doi, Nonaka,, and Behera}]{doi2020skill}
Doi, T., M.~Nonaka, and S.~Behera, 2020: {Skill Assessment of
  Seasonal-to-Interannual Prediction of Sea Level Anomaly in the North Pacific
  Based on the SINTEX-F Climate Model}. \textit{Front.\ Mar.\ Sci.},
  \textbf{7}, 546\,587.

\bibitem[{Dukowicz and Smith(1994)Dukowicz, and Smith}]{dukowicz1994implicit}
Dukowicz, J.~K., and R.~D. Smith, 1994: {Implicit free-surface method for the
  Bryan-Cox-Semtner ocean model}. \textit{J.\ Geophys.\ Res.\ Oceans},
  \textbf{99~(C4)}, 7991--8014.

\bibitem[{Dusek et~al.(2022)Dusek, Sweet, Widlansky, Thompson,, and
  Marra}]{dusek2022novel}
Dusek, G., W.~V. Sweet, M.~J. Widlansky, P.~R. Thompson, and J.~J. Marra, 2022:
  {A novel statistical approach to predict seasonal high tide flooding}.
  \textit{Front.\ Mar.\ Sci.}, \textbf{9}, 1073\,792.

\bibitem[{Eyring et~al.(2016)Eyring, Bony, Meehl, Senior, Stevens, Stouffer,,
  and Taylor}]{eyring2016overview}
Eyring, V., S.~Bony, G.~A. Meehl, C.~A. Senior, B.~Stevens, R.~J. Stouffer, and
  K.~E. Taylor, 2016: {Overview of the Coupled Model Intercomparison Project
  Phase 6 (CMIP6) experimental design and organization}. \textit{Geosci.\ Model
  Dev.}, \textbf{9~(5)}, 1937--1958.

\bibitem[{Eyring et~al.(2024)}]{eyring2024pushing}
Eyring, V., and Coauthors, 2024: Pushing the frontiers in climate modelling and
  analysis with machine learning. \textit{{Nat. Clim.\ Change}},
  \textbf{14~(9)}, 916--928.

\bibitem[{Fasullo et~al.(2020)Fasullo, Gent,, and Nerem}]{fasullo2020sea}
Fasullo, J.~T., P.~R. Gent, and R.~S. Nerem, 2020: {Sea Level Rise in the CESM
  Large Ensemble: The Role of Individual Climate Forcings and Consequences for
  the Coming Decades}. \textit{J.\ Climate}, \textbf{33~(16)}, 6911--6927.

\bibitem[{Feng et~al.(2025)}]{feng2025indications}
Feng, X., and Coauthors, 2025: {Indications of improved seasonal sea level
  forecasts for the United States Gulf and East Coasts using ocean-dynamic
  persistence}. \textit{EGUsphere Preprint repository}, \textbf{2025}, 1--23.

\bibitem[{Fox-Kemper et~al.(2021)}]{ipcc2021ocean}
Fox-Kemper, B., and Coauthors, 2021: {Climate Change 2021: The Physical Science
  Basis. Contribution of Working Group I to the Sixth Assessment Report of the
  Intergovernmental Panel on Climate Change}. Tech. rep., {Intergovernmental
  Panel on Climate Change}, {Cambridge, United Kingdom and New York, NY, USA}.

\bibitem[{Frame et~al.(2013)Frame, Methven, Gray,, and Ambaum}]{frame2013flow}
Frame, T., J.~Methven, S.~Gray, and M.~Ambaum, 2013: {Flow-dependent
  predictability of the North Atlantic jet}. \textit{Geophys.\ Res.\ Lett.},
  \textbf{40~(10)}, 2411--2416.

\bibitem[{Fraser et~al.(2019)Fraser, Palmer, Roberts, Wilson, Copsey,, and
  Zanna}]{fraser2019investigating}
Fraser, R., M.~Palmer, C.~Roberts, C.~Wilson, D.~Copsey, and L.~Zanna, 2019:
  {Investigating the predictability of North Atlantic sea surface height}.
  \textit{Climate Dyn.}, \textbf{53}, 2175--2195.

\bibitem[{Frederikse et~al.(2020)}]{frederikse2020causes}
Frederikse, T., and Coauthors, 2020: The causes of sea-level rise since 1900.
  \textit{Nature}, \textbf{584~(7821)}, 393--397.

\bibitem[{Fukumori et~al.(1998)Fukumori, Raghunath,, and
  Fu}]{fukumori1998nature}
Fukumori, I., R.~Raghunath, and L.-L. Fu, 1998: Nature of global large-scale
  sea level variability in relation to atmospheric forcing: A modeling study.
  \textit{J.\ Geophys.\ Res.\ Oceans}, \textbf{103~(C3)}, 5493--5512.

\bibitem[{Gill and Niller(1973)Gill, and Niller}]{gill1973theory}
Gill, A., and P.~Niller, 1973: The theory of the seasonal variability in the
  ocean. \textit{Deep-Sea Res. Oceanogr. Abstr.}, \textbf{20~(2)}, 141--177.

\bibitem[{Gneiting and Raftery(2007)Gneiting, and
  Raftery}]{gneiting2007strictly}
Gneiting, T., and A.~E. Raftery, 2007: Strictly proper scoring rules,
  prediction, and estimation. \textit{J.\ Am.\ Stat.\ Assoc.},
  \textbf{102~(477)}, 359--378.

\bibitem[{Gordon and Barnes(2022)Gordon, and Barnes}]{gordon2022incorporating}
Gordon, E.~M., and E.~A. Barnes, 2022: {Incorporating Uncertainty Into a
  Regression Neural Network Enables Identification of Decadal State-Dependent
  Predictability in CESM2}. \textit{Geophys.\ Res.\ Lett.}, \textbf{49~(15)},
  e2022GL098\,635.

\bibitem[{Gregory et~al.(2019)}]{gregory2019concepts}
Gregory, J.~M., and Coauthors, 2019: Concepts and terminology for sea level:
  Mean, variability and change, both local and global. \textit{Surv.\ Geophys},
  \textbf{40}, 1251--1289.

\bibitem[{Griffies et~al.(2014)}]{griffies2014assessment}
Griffies, S.~M., and Coauthors, 2014: {An assessment of global and regional sea
  level for years 1993--2007 in a suite of interannual CORE-II simulations}.
  \textit{Ocean Model.}, \textbf{78}, 35--89.

\bibitem[{Griffies et~al.(2016)}]{griffies2016omip}
Griffies, S.~M., and Coauthors, 2016: {OMIP contribution to CMIP6: Experimental
  and diagnostic protocol for the physical component of the Ocean Model
  Intercomparison Project}. \textit{Geosci.\ Model Dev.}, 3231.

\bibitem[{Guillaumin and Zanna(2021)Guillaumin, and
  Zanna}]{guillaumin2021stochastic}
Guillaumin, A.~P., and L.~Zanna, 2021: {Stochastic-Deep Learning
  Parameterization of Ocean Momentum Forcing}. \textit{J.\ Adv.\ Model.\ Earth\
  Syst.}, \textbf{13~(9)}, e2021MS002\,534.

\bibitem[{Ham et~al.(2019)Ham, Kim,, and Luo}]{ham2019deep}
Ham, Y.-G., J.-H. Kim, and J.-J. Luo, 2019: {Deep learning for multi-year ENSO
  forecasts}. \textit{Nature}, \textbf{573~(7775)}, 568--572.

\bibitem[{Hermans et~al.(2022)Hermans, Katsman, Camargo, Garner, Kopp,, and
  Slangen}]{hermans2022effect}
Hermans, T.~H., C.~A. Katsman, C.~M. Camargo, G.~G. Garner, R.~E. Kopp, and
  A.~B. Slangen, 2022: {The Effect of Wind Stress on Seasonal Sea-Level Change
  on the Northwestern European Shelf}. \textit{J.\ Climate}, \textbf{35~(6)},
  1745--1759.

\bibitem[{Hino et~al.(2019)Hino, Belanger, Field, Davies,, and
  Mach}]{hino2019high}
Hino, M., S.~T. Belanger, C.~B. Field, A.~R. Davies, and K.~J. Mach, 2019:
  High-tide flooding disrupts local economic activity. \textit{Sci.\ Adv.},
  \textbf{5~(2)}, eaau2736.

\bibitem[{Hochet et~al.(2024)Hochet, Llovel, Huck,, and
  S{\'e}vellec}]{hochet2024advection}
Hochet, A., W.~Llovel, T.~Huck, and F.~S{\'e}vellec, 2024: Advection
  surface-flux balance controls the seasonal steric sea level amplitude.
  \textit{Sci.\ Rep.}, \textbf{14~(1)}, 10\,644.

\bibitem[{Huber(1964)}]{huber1964robust}
Huber, P.~J., 1964: {Robust Estimation of a Location Parameter}. \textit{Ann.\
  Math.\ Stat.}, \textbf{35~(1)}, 73--101.

\bibitem[{Hummel et~al.(2018)Hummel, Berry,, and Stacey}]{hummel2018sea}
Hummel, M.~A., M.~S. Berry, and M.~T. Stacey, 2018: {Sea Level Rise Impacts on
  Wastewater Treatment Systems Along the U.S. Coasts}. \textit{{Earth's
  Future}}, \textbf{6~(4)}, 622--633.

\bibitem[{Immorlano et~al.(2025)Immorlano, Eyring, le~Monnier~de Gouville,
  Accarino, Elia, Mandt, Aloisio,, and Gentine}]{immorlano2025transferring}
Immorlano, F., V.~Eyring, T.~le~Monnier~de Gouville, G.~Accarino, D.~Elia,
  S.~Mandt, G.~Aloisio, and P.~Gentine, 2025: Transferring climate change
  physical knowledge. \textit{Proc.\ Natl.\ Acad.\ Sci.\ (USA)},
  \textbf{122~(15)}, e2413503\,122.

\bibitem[{Jacox et~al.(2020)}]{jacox2020seasonal}
Jacox, M.~G., and Coauthors, 2020: Seasonal-to-interannual prediction of north
  american coastal marine ecosystems: Forecast methods, mechanisms of
  predictability, and priority developments. \textit{Prog.\ Oceanogr.},
  \textbf{183}, 102\,307.

\bibitem[{Kalnay and Dalcher(1987)Kalnay, and Dalcher}]{kalnay1987forecasting}
Kalnay, E., and A.~Dalcher, 1987: Forecasting forecast skill. \textit{Mon.\
  Wea.\ Rev.}, \textbf{115~(2)}, 349--356.

\bibitem[{Kamp et~al.(2024)Kamp, Han, Zhang, Kido,, and
  McCreary}]{kamp2024tropical}
Kamp, W., W.~Han, L.~Zhang, S.~Kido, and J.~P. McCreary, 2024: {Tropical
  Atmospheric Intraseasonal Oscillations Leading to Sea Level Extremes in
  Coastal Indonesia during Recent Decades}. \textit{J.\ Climate},
  \textbf{37~(9)}, 2867--2880.

\bibitem[{Kenigson et~al.(2018)Kenigson, Han, Rajagopalan, Jasinski
  et~al.}]{kenigson2018decadal}
Kenigson, J.~S., W.~Han, B.~Rajagopalan, M.~Jasinski, and Coauthors, 2018:
  {Decadal Shift of NAO-Linked Interannual Sea Level Variability along the US
  Northeast Coast}. \textit{J.\ Climate}, \textbf{31~(13)}, 4981--4989.

\bibitem[{Kingma and Ba(2014)Kingma, and Ba}]{kingma2014adam}
Kingma, D.~P., and J.~Ba, 2014: {Adam: A Method for Stochastic Optimization}.
  \textit{arXiv preprint arXiv:1412.6980}, \doi{10.48550/arXiv.1412.6980}.

\bibitem[{Krishnamurthy(2019)}]{krishnamurthy2019predictability}
Krishnamurthy, V., 2019: Predictability of weather and climate. \textit{Earth
  Space Sci.}, \textbf{6~(7)}, 1043--1056.

\bibitem[{Krishnamurthy and Kirtman(2003)Krishnamurthy, and
  Kirtman}]{krishnamurthy2003variability}
Krishnamurthy, V., and B.~P. Kirtman, 2003: {Variability of the Indian Ocean:
  Relation to monsoon and ENSO}. \textit{Quart.\ J.\ Roy.\ Meteor.\ Soc.},
  \textbf{129~(590)}, 1623--1646.

\bibitem[{LeCun et~al.(2002)LeCun, Bottou, Orr,, and
  M{\"u}ller}]{lecun2002efficient}
LeCun, Y., L.~Bottou, G.~B. Orr, and K.-R. M{\"u}ller, 2002: {Efficient
  BackProp}. \textit{Neural Networks: Tricks of the Trade}, G.~M. et~al., Ed.,
  Springer, 9--48.

\bibitem[{Li et~al.(2022)}]{li2022contributions}
Li, S., and Coauthors, 2022: {Contributions of Different Sea-Level Processes to
  High-Tide Flooding Along the U.S. Coastline}. \textit{J.\ Geophys.\ Res.\
  Oceans}, \textbf{127~(7)}, e2021JC018\,276.

\bibitem[{Lin and Rood(1997)Lin, and Rood}]{lin1997explicit}
Lin, S.-J., and R.~B. Rood, 1997: {An explicit flux-form semi-Lagrangian
  shallow-water model on the sphere}. \textit{Quart.\ J.\ Roy.\ Meteor.\ Soc.},
  \textbf{123~(544)}, 2477--2498.

\bibitem[{Long et~al.(2021)}]{long2021seasonal}
Long, X., and Coauthors, 2021: {Seasonal Forecasting Skill of Sea-Level
  Anomalies in a Multi-Model Prediction Framework}. \textit{J.\ Geophys.\ Res.\
  Oceans}, \textbf{126~(6)}, e2020JC017\,060.

\bibitem[{Lorenz(1969)}]{lorenz1969predictability}
Lorenz, E.~N., 1969: The predictability of a flow which possesses many scales
  of motion. \textit{Tellus}, \textbf{21~(3)}, 289--307.

\bibitem[{Lorenz(1973)}]{lorenz1973existence}
Lorenz, E.~N., 1973: {On the Existence of Extended Range Predictability}.
  \textit{J.\ Appl.\ Meteor.}, 543--546.

\bibitem[{Mamalakis et~al.(2022)Mamalakis, Ebert-Uphoff,, and
  Barnes}]{mamalakis2022neural}
Mamalakis, A., I.~Ebert-Uphoff, and E.~A. Barnes, 2022: Neural network
  attribution methods for problems in geoscience: A novel synthetic benchmark
  dataset. \textit{Environmental Data Science}, \textbf{1}, e8.

\bibitem[{Mariotti et~al.(2018)Mariotti, Ruti,, and
  Rixen}]{mariotti2018progress}
Mariotti, A., P.~M. Ruti, and M.~Rixen, 2018: Progress in subseasonal to
  seasonal prediction through a joint weather and climate community effort.
  \textit{npj Climate Atmos.\ Sci.}, \textbf{1~(1)}, 4.

\bibitem[{Mariotti et~al.(2020)}]{mariotti2020windows}
Mariotti, A., and Coauthors, 2020: Windows of opportunity for skillful
  forecasts subseasonal to seasonal and beyond. \textit{Bull.\ Amer.\ Meteor.\
  Soc.}, \textbf{101~(5)}, E608--E625.

\bibitem[{Matheson and Winkler(1976)Matheson, and
  Winkler}]{matheson1976scoring}
Matheson, J.~E., and R.~L. Winkler, 1976: Scoring rules for continuous
  probability distributions. \textit{Manage. Sci.}, \textbf{22~(10)},
  1087--1096.

\bibitem[{Mayer and Barnes(2021)Mayer, and Barnes}]{mayer2021subseasonal}
Mayer, K.~J., and E.~A. Barnes, 2021: {Subseasonal Forecasts of Opportunity
  Identified by an Explainable Neural Network}. \textit{Geophys.\ Res.\ Lett.},
  \textbf{48~(10)}, e2020GL092\,092.

\bibitem[{Miles et~al.(2014)Miles, Spillman, Church,, and
  McIntosh}]{miles2014seasonal}
Miles, E.~R., C.~M. Spillman, J.~A. Church, and P.~C. McIntosh, 2014: Seasonal
  prediction of global sea level anomalies using an ocean-atmosphere dynamical
  model. \textit{Climate Dyn.}, \textbf{43}, 2131--2145.

\bibitem[{Murphy(1973)}]{murphy1973new}
Murphy, A.~H., 1973: {A new vector partition of the probability score}.
  \textit{J.\ Appl.\ Meteor.\ Climatol.}, \textbf{12~(4)}, 595--600.

\bibitem[{Murphy(1988)}]{murphy1988skill}
Murphy, A.~H., 1988: {Skill Scores Based on the Mean Square Error and Their
  Relationships to the Correlation Coefficient}. \textit{Mon.\ Wea.\ Rev.},
  \textbf{116~(12)}, 2417--2424.

\bibitem[{Nakkiran et~al.(2019)Nakkiran, Kaplun, Kalimeris, Yang, Edelman,
  Zhang,, and Barak}]{nakkiran2019sgd}
Nakkiran, P., G.~Kaplun, D.~Kalimeris, T.~Yang, B.~L. Edelman, F.~Zhang, and
  B.~Barak, 2019: {SGD on Neural Networks Learns Functions of Increasing
  Complexity}. \textit{Proceedings of the 33rd International Conference on
  Neural Information Processing Systems}, Curran Associates Inc., Red Hook, NY,
  USA.

\bibitem[{{NASEM}(2016)}]{nas2016next}
{NASEM}, 2016: \textit{Next Generation Earth System Prediction: Strategies for
  Subseasonal to Seasonal Forecasts}. The National Academies Press, Washington,
  DC, \doi{10.17226/21873}.

\bibitem[{Nix and Weigend(1994)Nix, and Weigend}]{nix1994estimating}
Nix, D.~A., and A.~S. Weigend, 1994: Estimating the mean and variance of the
  target probability distribution. \textit{{Proceedings of 1994 IEEE
  international conference on neural networks (ICNN'94)}}, IEEE, Vol.~1,
  55--60.

\bibitem[{O'Neill et~al.(2016)}]{oneill2016cmip6}
O'Neill, B.~C., and Coauthors, 2016: {The Scenario Model Intercomparison
  Project (ScenarioMIP) for CMIP6}. \textit{Geosci.\ Model Dev.},
  \textbf{9~(9)}, 3461--3482, \doi{10.5194/gmd-9-3461-2016},
  \urlprefix\url{https://gmd.copernicus.org/articles/9/3461/2016/}.

\bibitem[{Pan and Yang(2010)Pan, and Yang}]{pan2010survey}
Pan, S.~J., and Q.~Yang, 2010: A survey on transfer learning. \textit{{IEEE
  Trans.\ Knowl.\ Data Eng.}}, \textbf{22~(10)}, 1345--1359,
  \doi{10.1109/TKDE.2009.191}.

\bibitem[{Paszke et~al.(2019)}]{paszke2019pytorch}
Paszke, A., and Coauthors, 2019: Pytorch: an imperative style, high-performance
  deep learning library. \textit{Proceedings of the 33rd International
  Conference on Neural Information Processing Systems}, Curran Associates Inc.,
  Red Hook, NY, USA.

\bibitem[{Pedregosa et~al.(2011)}]{pedregosa2011sklearn}
Pedregosa, F., and Coauthors, 2011: {Scikit-learn: Machine Learning in Python}.
  \textit{J.\ Mach.\ Learn.\ Res.}, \textbf{12}, 2825--2830,
  \urlprefix\url{https://jmlr.csail.mit.edu/papers/v12/pedregosa11a.html}.

\bibitem[{Penduff et~al.(2010)Penduff, Juza, Brodeau, Smith, Barnier, Molines,
  Treguier,, and Madec}]{penduff2010impact}
Penduff, T., M.~Juza, L.~Brodeau, G.~C. Smith, B.~Barnier, J.-M. Molines, A.-M.
  Treguier, and G.~Madec, 2010: Impact of global ocean model resolution on
  sea-level variability with emphasis on interannual time scales. \textit{Ocean
  Sci.}, \textbf{6~(1)}, 269--284.

\bibitem[{Perezhogin et~al.(2023)Perezhogin, Zanna,, and
  Fernandez-Granda}]{perezhogin2023generative}
Perezhogin, P., L.~Zanna, and C.~Fernandez-Granda, 2023: {Generative
  Data-Driven Approaches for Stochastic Subgrid Parameterizations in an
  Idealized Ocean Model}. \textit{J.\ Adv.\ Model.\ Earth\ Syst.},
  \textbf{15~(10)}, e2023MS003\,681.

\bibitem[{Piecuch and Ponte(2011)Piecuch, and Ponte}]{piecuch2011mechanisms}
Piecuch, C., and R.~Ponte, 2011: Mechanisms of interannual steric sea level
  variability. \textit{Geophys.\ Res.\ Lett.}, \textbf{38~(15)}.

\bibitem[{Qiu(2002)}]{qiu2002large}
Qiu, B., 2002: {Large-scale variability in the midlatitude subtropical and
  subpolar North Pacific Ocean: Observations and causes}. \textit{J.\ Phys.\
  Oceanogr.}, \textbf{32~(1)}, 353--375.

\bibitem[{Qu et~al.(2022)Qu, Jevrejeva, Williams,, and Moore}]{qu2022drivers}
Qu, Y., S.~Jevrejeva, J.~Williams, and J.~C. Moore, 2022: {Drivers for seasonal
  variability in sea level around the China seas}. \textit{Global Planet.\
  Change}, \textbf{213}, 103\,819.

\bibitem[{Rao and Yamagata(2004)Rao, and Yamagata}]{rao2004abrupt}
Rao, S.~A., and T.~Yamagata, 2004: {Abrupt termination of Indian Ocean dipole
  events in response to intraseasonal disturbances}. \textit{Geophys.\ Res.\
  Lett.}, \textbf{31~(19)}.

\bibitem[{Roberts et~al.(2016)Roberts, Calvert, Dunstone, Hermanson, Palmer,,
  and Smith}]{roberts2016drivers}
Roberts, C., D.~Calvert, N.~Dunstone, L.~Hermanson, M.~Palmer, and D.~Smith,
  2016: {On the Drivers and Predictability of Seasonal-to-Interannual
  Variations in Regional Sea Level}. \textit{J.\ Climate}, \textbf{29~(21)},
  7565--7585.

\bibitem[{Rodgers et~al.(2021)}]{rodgers2021ubiquity}
Rodgers, K., and Coauthors, 2021: Ubiquity of human-induced changes in climate
  variability. \textit{Earth Syst.\ Dyn.}, \textbf{12~(4)}, 1393--1411.

\bibitem[{Rossby(1939)}]{rossby1939relation}
Rossby, C.-G., 1939: Relation between variations in the intensity of the zonal
  circulation of the atmosphere and the displacements of the semi-permanent
  centers of action. \textit{J.\ Mar.\ Res.}, \textbf{2}, 38--55.

\bibitem[{Saji et~al.(1999)Saji, Goswami, Vinayachandran,, and
  Yamagata}]{saji1999dipole}
Saji, N., B.~N. Goswami, P.~Vinayachandran, and T.~Yamagata, 1999: {A dipole
  mode in the tropical Indian Ocean}. \textit{Nature}, \textbf{401~(6751)},
  360--363.

\bibitem[{Salmon(1998)}]{salmon1998lectures}
Salmon, R., 1998: \textit{Lectures on Geophysical Fluid Dynamics}. OUP USA.

\bibitem[{Smith et~al.(2010)}]{smith2010parallel}
Smith, R., and Coauthors, 2010: {The Parallel Ocean Program (POP) Reference
  Manual}. \textit{LAUR-01853}, \textbf{141}, 1--140.

\bibitem[{Srivastava et~al.(2014)Srivastava, Hinton, Krizhevsky, Sutskever,,
  and Salakhutdinov}]{srivastava2014dropout}
Srivastava, N., G.~Hinton, A.~Krizhevsky, I.~Sutskever, and R.~Salakhutdinov,
  2014: {Dropout: A Simple Way to Prevent Neural Networks from Overfitting}.
  \textit{J.\ Mach.\ Learn.\ Res.}, \textbf{15~(1)}, 1929--1958.

\bibitem[{Stuecker et~al.(2017)Stuecker, Timmermann, Jin, Chikamoto, Zhang,
  Wittenberg, Widiasih,, and Zhao}]{stuecker2017revisiting}
Stuecker, M.~F., A.~Timmermann, F.-F. Jin, Y.~Chikamoto, W.~Zhang, A.~T.
  Wittenberg, E.~Widiasih, and S.~Zhao, 2017: {Revisiting ENSO/Indian Ocean
  dipole phase relationships}. \textit{Geophys.\ Res.\ Lett.}, \textbf{44~(5)},
  2481--2492.

\bibitem[{Sukop et~al.(2018)Sukop, Rogers, Guannel, Infanti,, and
  Hagemann}]{sukop2018high}
Sukop, M.~C., M.~Rogers, G.~Guannel, J.~M. Infanti, and K.~Hagemann, 2018:
  {High temporal resolution modeling of the impact of rain, tides, and sea
  level rise on water table flooding in the Arch Creek basin, Miami-Dade County
  Florida USA}. \textit{{Sci.\ Total Env.}}, \textbf{616}, 1668--1688.

\bibitem[{Sundararajan et~al.(2017)Sundararajan, Taly,, and
  Yan}]{sundararajan2017axiomatic}
Sundararajan, M., A.~Taly, and Q.~Yan, 2017: Axiomatic attribution for deep
  networks. \textit{International Conference on Machine Learning}, PMLR,
  3319--3328.

\bibitem[{Sweet and Park(2014)Sweet, and Park}]{sweet2014extreme}
Sweet, W.~V., and J.~Park, 2014: {From the extreme to the mean: Acceleration
  and tipping points of coastal inundation from sea level rise}.
  \textit{{Earth's Future}}, \textbf{2~(12)}, 579--600.

\bibitem[{Vallis(2017)}]{vallis2017atmospheric}
Vallis, G.~K., 2017: \textit{Atmospheric and Oceanic Fluid Dynamics}. Cambridge
  University Press.

\bibitem[{Vitart et~al.(2017)}]{vitart2017subseasonal}
Vitart, F., and Coauthors, 2017: {The Subseasonal to Seasonal (S2S) Prediction
  Project Database}. \textit{Bull.\ Amer.\ Meteor.\ Soc.}, \textbf{98~(1)},
  163--173.

\bibitem[{Wang et~al.(2023)Wang, Ren, Liu,, and Long}]{wang2023seasonal}
Wang, G., H.-L. Ren, J.~Liu, and X.~Long, 2023: {Seasonal predictions of sea
  surface height in BCC-CSM1.1m and their modulation by tropical climate
  dominant modes}. \textit{Atmos.\ Res.}, \textbf{281}, 106\,466.

\bibitem[{Wang et~al.(2016)Wang, Murtugudde,, and Kumar}]{wang2016evolution}
Wang, H., R.~Murtugudde, and A.~Kumar, 2016: {Evolution of Indian Ocean dipole
  and its forcing mechanisms in the absence of ENSO}. \textit{Climate
  Dynamics}, \textbf{47}, 2481--2500.

\bibitem[{Webster et~al.(1999)Webster, Moore, Loschnigg,, and
  Leben}]{webster1999coupled}
Webster, P.~J., A.~M. Moore, J.~P. Loschnigg, and R.~R. Leben, 1999: {Coupled
  ocean--atmosphere dynamics in the Indian Ocean during 1997--98}.
  \textit{Nature}, \textbf{401~(6751)}, 356--360.

\bibitem[{Wei et~al.(2021)Wei, Subramanian, Karnauskas, DeMott, Mazloff,, and
  Balmaseda}]{wei2021tropical}
Wei, H.-H., A.~C. Subramanian, K.~B. Karnauskas, C.~A. DeMott, M.~R. Mazloff,
  and M.~A. Balmaseda, 2021: {Tropical Pacific Air-Sea interaction processes
  and biases in CESM2 and their relation to El Ni{\~n}o development}.
  \textit{J.\ Geophys.\ Res.\ Oceans}, \textbf{126~(6)}, e2020JC016\,967.

\bibitem[{Widlansky et~al.(2014)Widlansky, Timmermann, McGregor, Stuecker,, and
  Cai}]{widlansky2014interhemispheric}
Widlansky, M.~J., A.~Timmermann, S.~McGregor, M.~F. Stuecker, and W.~Cai, 2014:
  {An interhemispheric tropical sea level seesaw due to El Ni{\~n}o Taimasa}.
  \textit{J.\ Climate}, \textbf{27~(3)}, 1070--1081.

\bibitem[{Widlansky et~al.(2017)}]{widlansky2017multimodel}
Widlansky, M.~J., and Coauthors, 2017: {Multimodel Ensemble Sea Level Forecasts
  for Tropical Pacific Islands}. \textit{J.\ Appl.\ Meteor.\ Climatol.},
  \textbf{56~(4)}, 849--862.

\bibitem[{Widlansky et~al.(2023)}]{widlansky2023quantifying}
Widlansky, M.~J., and Coauthors, 2023: {Quantifying the Benefits of Altimetry
  Assimilation in Seasonal Forecasts of the Upper Ocean}. \textit{J.\ Geophys.\
  Res.\ Oceans}, \textbf{128~(5)}, e2022JC019\,342.

\bibitem[{Wunsch and Stammer(1997)Wunsch, and Stammer}]{wunsch1997atmospheric}
Wunsch, C., and D.~Stammer, 1997: Atmospheric loading and the oceanic
  “inverted barometer” effect. \textit{Rev. Geophys.}, \textbf{35~(1)},
  79--107.

\bibitem[{Xie et~al.(2002)Xie, Annamalai, Schott,, and
  McCreary}]{xie2002structure}
Xie, S.-P., H.~Annamalai, F.~A. Schott, and J.~P. McCreary, 2002: {Structure
  and Mechanisms of South Indian Ocean Climate Variability}. \textit{J.\
  Climate}, \textbf{15~(8)}, 864--878.

\bibitem[{Yang et~al.(2015)Yang, Xie, Wu, Kosaka, Lau,, and
  Vecchi}]{yang2015seasonality}
Yang, Y., S.-P. Xie, L.~Wu, Y.~Kosaka, N.-C. Lau, and G.~A. Vecchi, 2015:
  {Seasonality and predictability of the Indian Ocean dipole mode: ENSO forcing
  and internal variability}. \textit{Journal of Climate}, \textbf{28~(20)},
  8021--8036.

\end{thebibliography}

\end{document}